\documentclass[showpacs,aps,prd,nofootinbib,floatfix,amsmath,amssymb]{iopart}
\usepackage{iopams}
\usepackage{graphicx}

\newcommand{\met}{E\!\!\!/_T }

\newcommand{\Pythia}{\mbox{{\sc pythia}}}
\newcommand{\Getjet}{\mbox{{\sc getjet}}}

\usepackage{epstopdf}
\begin{document}

\makeatletter
%Feynman slash
\newbox\slashbox \setbox\slashbox=\hbox{$/$}
\newbox\Slashbox \setbox\Slashbox=\hbox{\large$/$}
\def\pFMslash#1{\setbox\@tempboxa=\hbox{$#1$}
  \@tempdima=0.5\wd\slashbox \advance\@tempdima 0.5\wd\@tempboxa
  \copy\slashbox \kern-\@tempdima \box\@tempboxa}
\def\pFMSlash#1{\setbox\@tempboxa=\hbox{$#1$}
  \@tempdima=0.5\wd\Slashbox \advance\@tempdima 0.5\wd\@tempboxa
  \copy\Slashbox \kern-\@tempdima \box\@tempboxa}
\def\FMslash{\protect\pFMslash}
\def\FMSlash{\protect\pFMSlash}
\def\miss#1{\ifmmode{/\mkern-11mu #1}\else{${/\mkern-11mu #1}$}\fi}
%%%% Uso:  \pFMSlash{p}
\makeatother

\title{Higgs mediated Double Flavor Violating top decays in Effective Theories}
\author{A. Fern\' andez$^{(a)}$, C. Pagliarone$^{(b)}$, F. Ram\'\i rez-Zavaleta$^{(c)}$, and J. J. Toscano$^{(a)}$ }
\address{$^{(a)}$Facultad de Ciencias F\'{\i}sico Matem\'aticas,
Benem\'erita Universidad Aut\'onoma de Puebla, Apartado Postal
1152, Puebla, Pue., M\'exico.\\
$^{(b)}$Universit\'a di Cassino \& Istituto Nazionale di Fisica
Nucleare Pisa, Italy.\\
$^{(c)}$Facultad de Ciencias F\'\i sico Matem\' aticas,
Universidad Michoacana de San Nicol\' as de
Hidalgo, Avenida Francisco J. M\' ujica S/N, 58060, Morelia, Michoac\'an, M\' exico.}

%%%%%%%%%%%%%%%%%%%%%%%%%%%%%%%%%%%%%%%%%%%%%%%%%%%%%%%%%%%%%%%%%%%%%%%%%%%
\begin{abstract}
The possibility of detecting double flavor violating top quark
transitions at future colliders is explored in a
model-independent manner using the effective Lagrangian approach
through the $t \to u_i\tau \mu$ ($u_i=u,c$) decays. A Yukawa
sector that contemplates $SU_L(2)\times U_Y(1)$ invariants of up
to dimension six is proposed and used to derive the most general
flavor violating and CP violating $q_iq_jH$ and $l_il_jH$ vertices
of renormalizable type. Low-energy data, on high precision
measurements, and experimental limits are used to constraint the
$tu_iH$ and $H\tau \mu$ vertices and then used to predict the
branching ratios for the $t \to u_i\tau \mu$ decays. It is found
that this branching ratios may be of the order of $ 10^{-4}-10^{-5}$,
for a relative light Higgs boson with mass lower than $2m_W$,
which could be more important than those typical values found in
theories beyond the standard model for the rare top quark decays
$t\to u_iV_iV_j$ ($V_i=W,Z,\gamma, g$) or $t\to u_il^+l^-$. %%
LHC experiments, by using a total integrated luminosity of $\rm 3000~fb^{-1}$
of data, will be able to rule out, at 95\% C.L., DFV top quark decays up to a Higgs
mass of 155 GeV/$c^2$ or discover such a process up to a Higgs mass of $147$ GeV/$c^2$.
\end{abstract}
%%%%%%%%%%%%%%%%%%%%%%%%%%%%%%%%%%%%%%%%%%%%%%%%%%%%%%%%%%%%%%%%%%%%%%%%%%%


\maketitle

%%%%%%%%%%%%%%%%%%%%%%%%%%%%%%%%%%%%%%%%%%%%%%%%%%%%%%%%%%%%%%%%%%%%%%%%%%%
\section{Introduction}
Despite of the fact that the top quark is the heaviest known
particle, with a mass comparable to the electroweak symmetry
breaking scale, its dynamical behavior is rather restrictive. For
instance, its lifetime is so short that it decays before
hadronizing, almost exclusively, into the $bW$ mode. In fact, even
though the nondiagonal $t\to dW$ and $t\to sW$ decays have sizable
branching ratios, they play a quite marginal role compared with
the $t\to bW$ channel. For instance, the $t\to sW$ decay has a
branching ratio of order of $10^{-3}$. Rare three-body decays
generated at the tree level, as $t\to d_iWZ$ ($d_i=d,s$) and $t\to
u_iWW$ ($u_i=u,c$) are also severely restricted due to phase space
limitations and thus they strongly depend on the precise value of
the top mass~\cite{Elizabeth,TopTB}. At the one-loop level, the
interesting flavor changing neutral current (FCNC) decays $t\to
cV$ ($V=\gamma, g, Z, H$) are induced, but they have undetectable
branching ratios ranging from $10^{-10}$ to
$10^{-14}$~\cite{TopLoop,Mele}. These peculiarities, along with
diverse theoretical considerations, suggest that the top quark
could be very sensitive to new physics effects, which may manifest
themselves through anomalous rates for the top quark production
and decay modes. Although some properties of the top quark have
already been examined at the Tevatron~\cite{TopTevatron}, a
further scrutiny is expected at the CERN Large Hadron Collider
(LHC), which will operate as a veritable top quark factory,
producing about ten millions of $\bar{t}t$ events per year in
its first stage, and hopeful up to about eighteen millions in
subsequent years~\cite{TopLHC}. Some complementary studies will be
realized at $e^+e^-$ linear colliders~\cite{NLC}. Once
operating, many rare processes, involving this particle, are
expected to be accessible. It is thus worth investigating some
rare top quark decays that are very suppressed within the standard
model (SM), as they could constitute windows through which new
physics effects may show up.

In this paper, we are interested in the investigation of possible new
sources of flavor violation both in the quark and the lepton
sectors, through the double flavor violating top quark decay $t\to
u_i\tau^\pm \mu^\mp$\footnote{From now on, we will use the
notation $\tau \mu$ instead of $\tau^\pm \mu^\mp$.},
that could affect the SM top dilepton signatures by adding extra leptons
or by producing an asymmetry in the $ee$, $e \mu$, $\mu \mu$,
$e \tau$, $\mu \tau$ and $\tau \tau$ $+$ jets final states.
%%%%%%%%%%%%%%%%%%%%%%%%%%%%%%%%%%%%%%%%%%%%%%%%%%%%%%%%%%%%%%%%%%
Apart from its intrinsic interest, our motivation to study this rare process
was trigered from some top quark analysis published, by the CDF
Collaboration~\cite{ACDF} which revealed some level of asymmetry in the number
of $ee$, $\mu \mu$, and $e\mu$ $+$ jets events, coming from $t\bar{t}$ di-lepton
searches, that may be interpreted in terms of such types of rare top quark decays.
This decay is very suppressed within the SM, as it occurs at order $\alpha^4$
through a complicated sequence of subprocesses, namely, $t\to
Wb\to u_iWW \to u_i\tau \mu \nu_\tau \nu_\mu$. We will focus on
extended Yukawa sectors that are always present within the SM with
additional $SU_L(2)$-Higgs multiplets or in larger gauge groups.
Some processes naturally associated with flavor violation or CP
violation could be significantly impacted by Yukawa sectors
associated with multi-Higgs models, as it is expected that more
complicated Higgs sectors tend to favor this class of new physics
effects. We will assume that the double flavor violating decay
$t\to u_i\tau \mu$ is mediated by a virtual scalar field with a
mass of the order of the Fermi scale $v\approx 246$ GeV. However,
instead of tackling the problem in a specific model, we will adopt
a model independent approach by using the effective Lagrangian
technique~\cite{EL}, which is an appropriate scheme to study those
processes that are suppressed or forbidden in the SM. As it has been
shown in Refs.~\cite{T2,T3,PV}, it is not necessary to
introduce new degrees of freedom in order to generate flavor
violation at the level of classical action, the introduction of
operators of dimension higher than four will be enough. We will
see below that an effective Yukawa sector that incorporates
$SU_L(2)\times U_Y(1)$-invariants of up to dimension six is
enough to reproduce, in a model independent manner, the main
features that are common to extended Yukawa sectors, such as the
presence of both flavor violation and CP violation. Although
theories beyond the SM require more complicated Higgs sectors that
include new physical scalars, we stress that our approach for
studying the $t\to u_i\tau \mu$ decay mediated by a relatively
light scalar particle is sufficiently general to incorporate the
most relevant aspects of extended theories, as in most cases, it
is always possible to identify in an appropriate limit a SM-like
Higgs boson whose couplings to pairs of $W$ and $Z$ bosons
coincide with those given in the minimal SM. Besides its model
independence, our framework has the advantage that it involves an
equal or even less number of unknown parameters than those usually
appearing in specific extended Yukawa sectors. One important
ingredient of our study consists in supporting our predictions, as
much as possible, in the available experimental data. For this
purpose, in order to predict the branching ratio of the $t\to
u_i\tau \mu$ decay, we will use the current low-energy data to
get bounds on the flavor violating $Htu_i$ and $H\tau \mu$
couplings. As we will see below, the best constraint on the
$Htu_i$ coupling arises from the recently observed
$D^0-\overline{D^0}$ mixing~\cite{BaBar,Belle,HFAG}, although the
experimental uncertainty on the proton dipole moment provides a
lightly lower bound. As to the $H\tau \mu$ coupling, we will see
that the best bound arises from the experimental uncertainty on
the muon magnetic dipole moment. A less stringent bound for the $H\tau
\mu$ coupling is derived from the experimental limit on the
branching ratio of the $\tau \to \mu \bar{\mu}\mu$ decay. We will
present explicit expressions that can eventually be used for
bounding the parameters of any specific model. Our expression for
the branching ratio of the $t\to u_i\tau \mu$ decay is also of
general applicability, as it may be adapted easily to any specific
model.

The paper has been organized as follows. In Section~\ref{y} the gauge
structure of an effective Yukawa sector that include up to
dimension-six $SU_L(2)\times U_Y(1)$-invariant operators is
discussed. Section~\ref{d} is devoted to calculate the branching
ratio for the $t\to u_i\tau \mu$ decay. In Section~\ref{b} a
comprehensive analysis concerning the impact of the effective
Yukawa sector on low-energy observables is carry out in order to
determine the best constraints for the $Htu_i$ and $H\tau \mu$
flavor violating couplings. Section~\ref{r} is reserved to present
results and analyze the experimental perspectives. Finally, in
Section~\ref{c} the conclusions are presented.

%%%%%%%%%%%%%%%%%%%%%%%%%%%%%%%%%%%%%%%%%%%%%%%%%%%%%%%%%%%%%%%%%%%%%%%%%%% EFFECTIVE LAGRANGIAN
\section{Effective Lagrangian for the Yukawa sector}
\label{y} In the SM the Yukawa sector is both flavor-conserving and
CP-conserving, but these effects can be generated at the tree
level if new scalar fields are introduced. One alternative, which
does not contemplate the introduction of new degrees of freedom,
consists in incorporating into the classical action the virtual
effects of the heavy degrees by introducing $SU_L(2)\times
U_Y(1)$-invariant operators of dimension higher than four~\cite{T2,T3}.
Indeed, it is only necessary to extend the Yukawa
sector with dimension-six operators to induce the most general
coupling of the Higgs boson to quarks and leptons. A Yukawa sector
with these features has the following structure~\cite{T2,T3}
\begin{eqnarray}
{\cal L}^Y_{eff}&=&-Y^l_{ij}(\bar{L}_i\Phi
l_j)-\frac{\alpha^i_{ij}}{\Lambda^2}(\Phi^\dag \Phi)(\bar{L}_i\Phi
l_j)+ H.c. \nonumber \\
&&-Y^d_{ij}(\bar{Q}_i\Phi
d_j)-\frac{\alpha^d_{ij}}{\Lambda^2}(\Phi^\dag \Phi)(\bar{Q}_i\Phi
d_j)+ H.c.\\
&&-Y^u_{ij}(\bar{Q}_i\tilde{\Phi}
u_j)-\frac{\alpha^u_{ij}}{\Lambda^2}(\Phi^\dag
\Phi)(\bar{Q}_i\tilde{\Phi} u_j)+ H.c.\nonumber,
\end{eqnarray}
where $Y_{ij}$, $L_i$, $Q_i$, $\Phi$, $l_i$, $d_i$, and $u_i$
stand for the usual components of the Yukawa matrix, the
left-handed lepton doublet, the left-handed quark doublet, the
Higgs doublet, the right-handed charged lepton singlet, and the
right-handed quark singlets of down and up type, respectively.
The $\alpha_{ij}$ numbers are the components of a $3\times 3$
general matrix, which parametrize the details of the underlying
physics, whereas $\Lambda$ is the typical scale of these new
physics effects.

After spontaneous symmetry breaking, this extended Yukawa sector
can be diagonalized as usual via the unitary matrices
$V^{l,d,u}_L$ and $V^{l,d,u}_R$, which correlate gauge states to mass
eigenstates. In the unitary gauge, the diagonalized Lagrangian can
be written as follows:
\begin{eqnarray}
{\cal
L}^Y_{eff}&=&-\Big(1+\frac{g}{2m_W}H\Big)\Big(\bar{E}M_lE+\bar{D}M_dD+\bar{U}M_uU\Big)\nonumber
\\
&&-H\Big(1+\frac{g}{4m_W}H\Big(3+\frac{g}{2m_W}H\Big)\Big)\Big(\bar{E}\Omega^lP_RE+\bar{D}\Omega^d
P_RD\nonumber\\
&&+\bar{U}\Omega^u P_RU+H.c.\Big),
\end{eqnarray}
where the $M_a$ ($a=l,d,u$) are the diagonal mass matrix and
$\bar{E}=(\bar{e},\bar{\mu},\bar{\tau})$,
$\bar{D}=(\bar{d},\bar{s},\bar{b})$, and
$\bar{U}=(\bar{u},\bar{c},\bar{t})$ are vectors in the flavor
space. In addition, $\Omega^a$ are matrices defined in the flavor
space through the relation:
\begin{equation}
\Omega^a=\frac{1}{\sqrt{2}}\Big(\frac{v}{\Lambda}\Big)^2V^a_L\alpha^aV^{a\dag}_R.
\end{equation}
To generate Higgs-mediated FCNC effects at the level of classical
action, it is assumed that neither $Y^{l,d,u}$ nor
$\alpha^{l,d,u}$ are diagonalized by the $V^a_{L,R}$ rotation
matrices, which should only diagonalize the sum
$Y^{l,d,u}+\alpha^{l,d,u}$. As a consequence, mass and
interactions terms would not be simultaneously diagonalized as it
occurs in the dimension-four theory. In addition, if
$\Omega^{a\dag}\neq \Omega^a$, the Higgs boson couples to fermions
through both scalar and pseudoscalar components, which in turn
could lead to CP violation in some processes. As a consequence,
the flavor violating coupling $\bar{f}_if_jH$, with $f$ stands for
a charged lepton~\cite{HLFV} or quark, has the most general renormalizable
structure of scalar and pseudoscalar type given by~\footnote{The
flavor violating coupling $\bar{f}_if_jH$ can also be induced by
 $\overline{D_\mu F}_if_{Rj}D^\mu \Phi$ operators, with $D_\mu$ the electroweak covariant derivative and
$F$ and $f$ stand for doublet and singlet, respectively. However, they would have a marginal role in flavor physics, as
this class of operators only may be generated at the one-loop
level by the fundamental theory and are thus suppressed with
respect to the Yukawa type ones, which can be generated at the
tree level~\cite{AEW}. In addition, these operators induce also
the $Zf_if_j$ coupling, allowing us to constraint it through
$Z$ physics data~\cite{T3}.}:
\begin{equation}
-i(\Omega_{ij}P_R+\Omega^*_{ij}P_L)=-i[Re(\Omega_{ij})+iIm(\Omega_{ij})\gamma_5].
\end{equation}

To close this section, let us to emphasize that the above
effective Lagrangian describes
 %the most general coupling of renormalizable type of a scalar field to pairs of fermions,
 the most general renormalizable coupling of scalar field to pairs of fermions,
 which reproduces the main features of most of the extended Yukawa sectors. It includes
the most general version of the two-Higgs doublet model
(THDM-III)~\cite{THDM-III} and  multi-Higgs models that comprise
additional multiplets of $SU_L(2)\times U_Y(1)$ or scalar
representations of larger gauge groups. Our approach also cover
more exotic formulations of flavor violation, as the so-called
familons models~\cite{Familons} or theories that involves an
Abelian flavor symmetry~\cite{AFS}. In this way, our results will
be applicable to a wide variety of models that predict
scalar-mediated flavor violation.
%%%%%%%%%%%%%%%%%%%%%%%%%%%%%%%%%%%%%%%%%%%%%%%%%%%%%%%%%%%%%%%%%%%%%%%%%%%

%%%%%%%%%%%%%%%%%%%%%%%%%%%%%%%%%%%%%%%%%%%%%%%%%%%%%%%%%%%%%%%%%%%%%%%%%%% THE DECAY top...
\section{The decay $t\to u_i\tau \mu$}
\label{d} 
\begin{figure}
\centering
\includegraphics[width=3.0in]{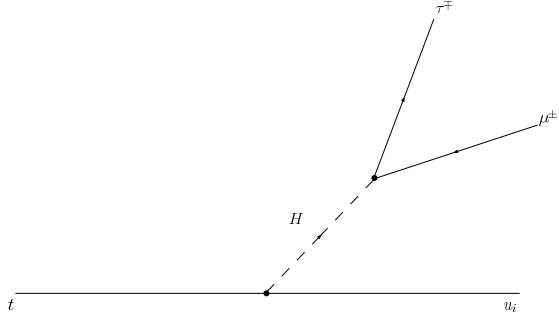}
\caption{\label{TD} Diagram for double flavor violating top quark
decay.}
\end{figure}

We now proceed to calculate the branching ratio for the
$t\to u_i\tau \mu$ decay using the general couplings for $tu_iH$
and $H\tau \mu$ given in the previous section. This decay occurs
through the diagram shown in Figure~\ref{TD}. The amplitude can be
written as the product of the amplitudes associated with the
subprocesses $t\to u_iH$ and $H\to \tau \mu$:
\begin{equation}
{\cal M}(t\to u_i\tau \mu)=\frac{{\cal M}(t\to u_iH){\cal M}(H\to
\tau \mu)}{m^2-m^2_H+im_H\Gamma_H},
\end{equation}
where $m$ is the invariant mass of the final state $\tau \mu$ and
$\Gamma_H$ is the total Higgs decay width. Once squared the
amplitude, averaging over the top spin states, and introducing a
factor of $2$ to take into account the two final states
$\tau^+\mu^-$ and $\tau^-\mu^+$, one obtains
\begin{equation}
|\bar{{\cal M}}(t\to u_i\tau \mu)|^2=\frac{|\bar{{\cal M}}(t\to
u_iH)|^2|{\cal M}(H\to \tau \mu^)|^2}{m^4_t((x-y)^2+yz)},
\end{equation}
where we have introduced the dimensionless variables
$x=m^2/m^2_t$, $y=m^2_H/m^2_t$, and $z=\Gamma^2_H/m^2_t$. In
addition, the squared subamplitudes are given by
\begin{eqnarray}
|\bar{{\cal M}}(t\to
u_iH)|^2&=&m^2_t\Big((1-x)|\Omega_{u_it}|^2+4\Big(\frac{m_{u_i}}{m_t}\Big)Re(\Omega^2_{u_it})\Big),\\
|{\cal M}(H\to \tau \mu)|^2&=&m^2_t\Big(x|\Omega_{\tau
\mu}|^2+4\Big(\frac{m_\tau}{m_t}\Big)\Big(\frac{m_\mu}{m_t}\Big)Re(\Omega^2_{\tau
\mu})\Big).
\end{eqnarray}
From now on, the masses $m_{u_i}$ and $m_\mu$ will be ignored with respect to $m_t$ and $m_H$, in
phase space manipulations.
 In this approach, the branching ratio for this decay can be written as
follows:
\begin{equation}
Br(t\to u_i\tau \mu)=\frac{|\Omega_{u_it}|^2|\Omega_{\tau
\mu}|^2}{256\pi^3}\Big(\frac{m_t}{\Gamma_t}\Big)f(y,z),
\end{equation}
where $\Gamma_t$ is the top total decay width and
\begin{equation}
f(y,z)=\int^1_0 dx\frac{x(1-x)^2}{(x-y)^2+yz}.
\end{equation}
This integral has analytical solution, which is given by
\begin{eqnarray}
\label{ff}
f(y,z)&=&2y-\frac{3}{2}+\frac{1}{2}\Big(3y^2-(4+z)y+1\Big)\log\Bigg(\frac{y^2+(z-2)y+1}{y(y+z)}\Bigg)\nonumber
\\
&&+\sqrt{\frac{y}{z}}\Big(y^2-(2+3z)y+2z+1\Big)\Bigg(\tan^{-1}\sqrt{\frac{y}{z}}\nonumber\\
&&+\tan^{-1}\Big(\frac{1-y}{\sqrt{yz}}\Big)\Bigg).
\end{eqnarray}

In order to make predictions, we need to use some values for the
$\Omega_{u_it}$ and $\Omega_{\tau \mu}$ parameters. This is the
subject of the next section.

%%%%%%%%%%%%%%%%%%%%%%%%%%%%%%%%%%%%%%%%%%%%%%%%%%%%%%%%%%%%FIGURE 1
%%%%%%%%%%%%%%%%%%%%%%%%%%%%%%%%%%%%%%%%%%%%%%%%%%%%%%%%%%%%%%%%%%%

%%%%%%%%%%%%%%%%%%%%%%%%%%%%%%%%%%%%%%%%%%%%%%%%%%%%%%%%%%%%%%%%%%%%%%%%%%% BOUNDING THE H...
\section{Bounding the $Htu_i$ and $H\tau \mu$ vertices
from low-energy data} \label{b} Our goal in this section is to
estimate the size of the $\Omega_{u_it}$ and $\Omega_{\tau \mu}$
parameters. Although they have already been estimated before by
means of the Cheng-Sher ansatz~\cite{SHER1} or other similar
assumptions~\cite{T2}, in this work, as emphasized in the
introduction, we will resort to the available low-energy data. It
is obvious that a constraint on the $\Omega_{u_it}$ parameter from
low-energy data only can be obtained through one-loop or higher
order effects. In this case, we will examine the simultaneous
one-loop contribution of the $Htu$ and $Htc$ vertices to the mass
difference $\Delta M_D$ in $D^0-\overline{D^0}$ mixing recently
observed by the Babar~\cite{BaBar} and Belle~\cite{Belle}
collaborations. Also, the one-loop contribution of the $Htu$
vertex to the magnetic and electric dipole moments of the proton
and the neutron will be examined. The contribution of the $Htc$
vertex to these observables would require to consider two-loop
effects. However, as we will discuss below, the constraint thus
obtained is not appropriated to make predictions. As far as the
$\Omega_{\tau \mu}$ parameter is concerned, there are two possible
type of processes that eventually could lead to good constraints.
These processes are the experimental uncertainty on the muon
anomalous magnetic moment and the experimental limits on the
lepton flavor violating decays $l_i \to l_j \bar{l}_kl_k$ and
$l_i \to l_j \gamma$. We will see below that in this case the
experimental uncertainty on the muon anomalous magnetic moment
provides the best constraint.

\subsection{Bounding $\Omega_{tu}\Omega_{tc}$ from
$D^0-\overline{D^0}$ mixing} 
\begin{figure}
\centering
\includegraphics[width=3.5in]{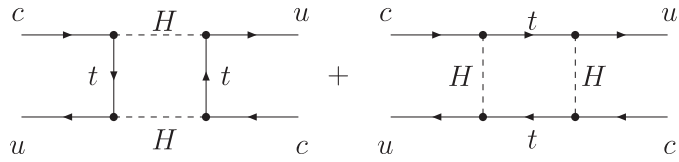}
\caption{\label{DD} Box diagrams contributing to the $D^0-\overline{D^0}$ mixing.}
\end{figure}

Recently, the Babar~\cite{BaBar} and
Belle~\cite{Belle} collaborations have observed signals of
$D^0-\overline{D^0}$ mixing. The Heavy Flavor Averaging
Group~\cite{HFAG} interpretation of the current data leads to a
mass difference of $D^0-\overline{D^0}$ given by $\Delta
M_D=(1.4\pm 0.5)\times 10^{-15}$ GeV$/c^2$. Although this
measurement is recent and fluctuations from future refinements can
be expected, analysis carried out show that this value is
consistent with the SM prediction~\cite{GHPP}. In
Ref.~\cite{GHPP}, diverse scenarios of new physics contributions
to $D$ mixing are studied. Here, we will
investigate the impact of the $Htc$ and $Htu$ vertices on the mass
difference  $\Delta M_D$. This contributions occurs through the
short distance effect characterized by the box diagrams shown
in Fig.\ref{DD}. The amplitude is entirely governed by the top and Higgs masses, so we will work in the limit of external masses and momenta equal to zero. In this approximation, the amplitude can  be written as follows
\begin{equation}
\Gamma=2\Omega^2_{tu}\Omega^2_{tc}\int \frac{d^4k}{(2\pi)^4}\frac{(\bar{u}(\pFMSlash{k}+m_t)c)(\bar{u}(\pFMSlash{k}+m_t)c)}{(k^2-m^2_t)^2(k^2-m_H)^2},
\end{equation}
where the $\Omega_i$ parameters have been assumed reals. Notice that the amplitude is free of ultraviolet divergences. Once introduced a Feynman parametrization, the amplitude can be written in the following form
\begin{equation}
\Gamma=-\frac{i}{16\pi^2}\frac{\Omega^2_{tu}\Omega^2_{tc}}{m^2_t}\Big(f(x)(\bar{u}\gamma_\mu c)(\bar{u}\gamma^\mu c)+
g(x)(\bar{u}c)(\bar{u}c)\Big),
\end{equation}
where $f(x)$ and $g(x)$ ($x=m^2_H/m^2_t$) are the loop functions in the heavy internal mass limit, which are given by
\begin{equation}
f(x)=\frac{1}{2}\frac{1}{(1-x)^3}(1-x^2+2x\log x),
\end{equation}
\begin{equation}
g(x)=\frac{4}{(1-x)^3}\Big(2(1-x)+(1+x)\log x\Big),
\end{equation}
where $f(1)=1/6$ and $g(1)=-2/3$. The $\Gamma$ amplitude corresponds to the vertex function
associated with the following four-quark effective interaction:
\begin{equation}
{\cal
L}_{eff}=-\frac{\Omega^2_{tu}\Omega^2_{tc}}{64\pi^2 m^2_t}\Big(f(x)(Q_1+2Q_2+Q_6)+g(x)(Q_3+2Q_4+Q_7)\Big),
\end{equation}
where a factor of $1/4$ was introduced in order to compensate two Wick contractions. Here, the $Q_{i}$ are dimension-six fermionic operators, which are already known in the literature~\cite{GHPP}:
\begin{eqnarray}
Q_1&=&(\bar{u}_L\gamma_\mu c_L)(\bar{u}_L\gamma^\mu c_L), \\
Q_2&=&(\bar{u}_L\gamma_\mu c_L)(\bar{u}_R\gamma^\mu c_R),\\
Q_6&=&(\bar{u}_R\gamma_\mu c_R)(\bar{u}_R\gamma^\mu c_R),\\
Q_3&=&(\overline{u}_Lc_R)(\overline{u}_Rc_L),\\
Q_4&=&(\overline{u}_Rc_L)(\overline{u}_Rc_L),\\
Q_7&=&(\overline{u}_Lc_R)(\overline{u}_Lc_R).
\end{eqnarray}

\begin{figure}
\centering
\includegraphics[width=3.5in]{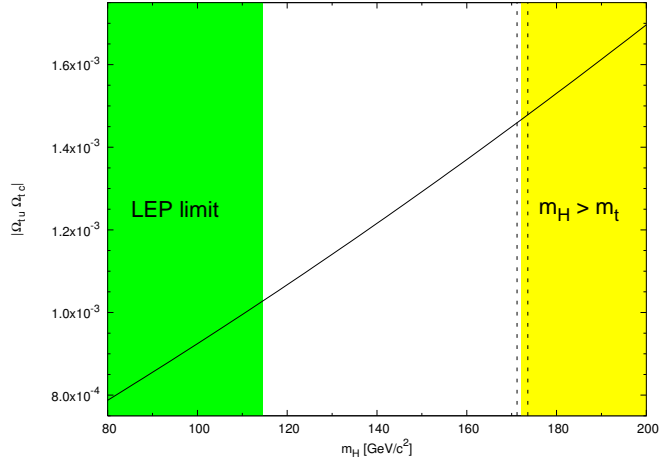}
\caption{\label{otcotu} Behavior of $\Omega_{tc}\Omega_{tu}$ with the Higgs mass. The
dashed lines represent the current experimental uncertainty on the top quark mass measurement~\cite{PDG}.}
\end{figure}

On the other hand, the $D$ mass difference is given by
\begin{equation}
\Delta M_D=\frac{1}{M_D}Re<\overline{D^0}|{\cal H}_{eff}=-{\cal
L}_{eff}|D^0>,
\end{equation}
which in our case takes the form
\begin{eqnarray}
\Delta
M_D&=&\frac{\Omega^2_{tu}\Omega^2_{tc}}{64\pi^2}\frac{1}{m^2_tM_D}\Big(f(x)<Q_1>+2<Q_2>+<Q_6>)\nonumber\\
&&+g(x)(<Q_3>+2<Q_4>+<Q_7>)\Big)\nonumber
\\
&=&\frac{\Omega^2_{tu}\Omega^2_{tc}}{64\pi^2}\Big(\frac{f_D}{m_t}\Big)^2\Big(\frac{M_D}{12}\Big)\Big(f(x)(8B_1-20B_2+8B_6)\nonumber\\
&&+g(x)(7B_3-10B_4-5B_7)\Big),
\end{eqnarray}
where the expressions for $<Q_i>$ given in Ref.~\cite{GHPP} were
used. Here, $f_D$ is the $D$ decay constant. We will use the CLEO
Collaboration determination $f_D=222.6 \pm 16.7$
MeV$/c^2$~\cite{CLEOfd}. The factors $B_i$ are unknown, but lattice
calculation~\cite{Lattice} leads to $B_i\approx 0.8$~\cite{GHPP},
although in vacuum saturation and in the heavy quark limit they
approximate to the unity. In our numerical analysis, we will use
$B_i=1$. Using the value $M_D=1.8646$ GeV$/c^2$~\cite{PDG}, one obtains
\begin{equation}
\Delta M_D=1.6\times 10^{-9}\Big(f(x)-2g(x)\Big)\Omega^2_{tu}\Omega^2_{tc}.
\end{equation}
To bound the $\Omega_{tu}\Omega_{tc}$ product, we demand that the
above contribution does not exceed the experimental
uncertainty~\cite{BaBar,Belle,HFAG}, which leads to
\begin{equation}
\label{CDD}
|\Omega_{tu}\Omega_{tc}|<\frac{1.8\times 10^{-3}}{\sqrt{F(x)}},
\end{equation}
where $F(x)=f(x)-2g(x)$. In Figure \ref{otcotu} we show the behavior of
$|\Omega_{tu}\Omega_{tc}|$ as a function of the Higgs mass. From
this figure, it can be appreciated a moderate variation of
$|\Omega_{tu}\Omega_{tc}|$, which go from $0.8\times 10^{-3}$ to
$1.6\times 10^{-3}$ for $m_H$ ranging from $80$ GeV$/c^2$ to $200$ GeV$/c^2$.

%%%%%%%%%%%%%%%%%%%%%%%%%%%%%%%%%%%%%%%%%%%%%%%%%%%%%%%%%%%%FIGURE 1
%%%%%%%%%%%%%%%%%%%%%%%%%%%%%%%%%%%%%%%%%%%%%%%%%%%%%%%%%%%%%%%%%%%

%%%%%%%%%%%%%%%%%%%%%%%%%%%%%%%%%%%%%%%%%%%%%%%%%%%%%%%%%%%%FIGURE 1
%%%%%%%%%%%%%%%%%%%%%%%%%%%%%%%%%%%%%%%%%%%%%%%%%%%%%%%%%%%%%%%%%%%

\subsection{Bounds from flavor diagonal low-energy processes}
In this paragraph we will calculate the contribution of the
$Hf_if_j$ vertex to the magnetic and electric dipole moments of the
$f_i$ fermion, which is given through the diagram shown in
Figure~\ref{BT}. In order to have some idea about the importance of
the imaginary part of the $\Omega_{ij}$ parameter, we will proceed
as general as possible, which means that the contribution to both
the magnetic and electric dipole moments of $f_i$ will be
calculated. Here, the $f_i$ fermion stands for a light quark or
charged lepton. The amplitude for the on-shell $f_if_j\gamma$
vertex can be written as follows:
\begin{equation}
{\cal M}=\bar{u}(p_2)\Gamma_\mu u(p_1)\epsilon^{*\mu}(q,\lambda),
\end{equation}
where the vertex function is given by
\begin{equation}
\Gamma_\mu=eQ_j\int\frac{d^Dk}{(2\pi)^D}\frac{T_\mu}{\Delta},
\end{equation}
with
\begin{eqnarray}
T_\mu&=&|\Omega_{ij}|^2\Big((\pFMSlash{k}+\pFMSlash{p_2})\gamma_\mu(\pFMSlash{k}+\pFMSlash{p_1})+m^2_j\gamma_\mu\Big)+\nonumber
\\
&&m_j\Big(Re(\Omega^2_{ij})+iIm(\Omega^{*2}_{ij})\gamma_5\Big)(2k_\mu
+\pFMSlash{p_2}\gamma_\mu+\gamma_\mu \pFMSlash{p_1}),
\end{eqnarray}

\begin{equation}
\Delta=[k^2-m^2_H][(k+p_1)^2-m^2_j][(k+p_2)^2-m^2_j].
\end{equation}
In the above expressions, $Q_j$ is the electric charge of the
internal $f_j$ fermion in units of the positron charge. After using
the well known Gordon's identities, it is easy to see that there are
contributions to the monopole [$F(q^2)\gamma_\mu$], the magnetic
dipole moment [$i(a_i/2m_i)\sigma_{\mu \nu}q^\nu$], and the electric
dipole moment ($-d_i\gamma_5\sigma_{\mu \nu}q^\nu$). The
contribution to the monopole is divergent, but we only are
interested in the magnetic and electric dipole moments, for which
the contribution is free of ultraviolet divergences. After some
algebra, the form factors associated with the electromagnetic
dipoles of $f_i$ can be written as follows:
\begin{equation}
\label{md}
a_i=-\frac{Q_jx^2_i}{8\pi^2}\left(|\Omega_{ij}|^2f(x_i,x_j)+\Big(\frac{x_j}{x_i}\Big)Re(\Omega^2_{ij})g(x_i,x_j)\right),
\end{equation}

\begin{equation}
\label{ed}
d_i=-\frac{Q_jx_j}{16\pi^2}\Big(\frac{e}{m_H}\Big)Im(\Omega^2_{ij})g(x_i,x_j),
\end{equation}
where
\begin{eqnarray}
f(x_i,x_j)&=&\int^1_0dx\int^{1-x}_0dy\frac{(1-x-y)(x+y)}{R},\\
g(x_i,x_j)&=&\int^1_0dx\int^{1-x}_0dy\frac{(x+y)}{R},
\end{eqnarray}
with
\begin{equation}
R=1-x-y-x^2_i(1-x-y)(x+y)+x^2_j(x+y).
\end{equation}
In the above expressions the dimensionless variable $x_a=m_a/m_H$
has been introduced. The solutions of these integrals are quite
complicated, but we can take $x_i=0$ in $R$ to obtain simple
analytical expressions, which is possible since $x_i\ll x_j$ for all
the cases that will be considered here, namely, $m_i=m_u,m_d,m_\mu$.
So in this approximation, one obtains
\begin{eqnarray}
f(x_j)&=&\frac{2+x^2_j(x^4_j-6x^2_j+3)+6x^2_jlog(x^2_j)}{6(x^2_j-1)^4},\\
g(x_j)&=&\frac{3+x^2_j(x^2_j-4)+2log(x^2_j)}{2(x^2_j-1)^3}.
\end{eqnarray}
A simple numerical evaluation shows that $f(x_j)$ is suppressed with
respect to $g(x_j)$ by about one order of magnitude. Consequently,
the $x^2_if(x_j)$ term can be neglected in the expression for $a_i$,
as it is irrelevant compared with $x_ix_jg(x_j)$. After these
considerations, we can write
\begin{eqnarray}
a_i&=&-\frac{Q_jx_ix_j}{8\pi^2}Re(\Omega^2_{ij})g(x_j),\\
d_i&=&-\frac{Q_jx_j}{16\pi^2}\Big(\frac{e}{m_H}\Big)Im(\Omega^2_{ij})g(x_j).
\end{eqnarray}
Notice that $Re(\Omega^2_{ij})=[Re(\Omega_{ij})]^2-[Im(\Omega_{ij})]^2$, which means that the scalar (pseudoscalar) component of the $Hf_if_j$ vertex gives a  positive (negative) contribution to the magnetic dipole moment of the fermion in consideration, a fact that is well known from the abundant literature on the muon's anomalous magnetic moment.

We now are in position of using low-energy data to bound the
$\Omega_{tu}$ and $\Omega_{\tau \mu}$ parameters.

\subsubsection{Bounding $\Omega_{tu}$ from proton and neutron
physics}

To bound $Re(\Omega^2_{tu})$ we can use the available precision
measurements on the magnetic dipole moments of the proton and the
neutron. We will use the experimental uncertainty on the proton
magnetic dipole, as it is the most restrictive. Thus, the
contribution of the $tuH$ vertex must be less than this uncertainty,
which is given by~\cite{PDG}
\begin{equation}
|\Delta a^{Exp}_p|< 2.8 \times 10^{-8}.
\end{equation}
As far as $Im(\Omega^2_{tu})$ is concerned, we can constraint it
using the very stringent experimental limit for the neutron electric
dipole moment. The current limit is given by~\cite{EDMN}:
\begin{equation}
|d_n|<2.9\times 10^{-26}\ e.cm.
\end{equation}

\begin{figure}
\centering
\includegraphics[width=4.0in]{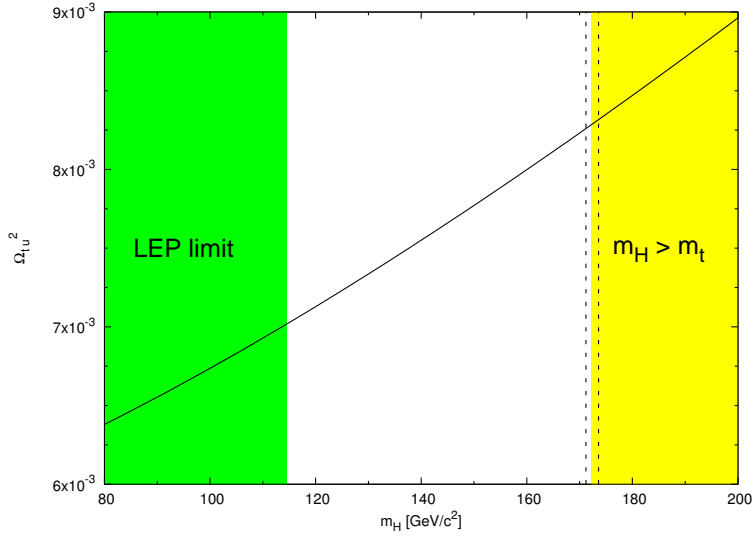}
\caption{\label{otu}Behavior of $\Omega^2_{tu}$ as a function of
the Higgs mass.}
\end{figure}

On the other hand, the magnetic and electric dipole moments of $p$
and $n$, respectively, are related with those of their elementary
constituents through the following expressions
\begin{eqnarray}
a_p&=&\frac{2}{3}a_u-\frac{1}{3}a_d-\frac{1}{3}a_s,\\
d_n&=&\frac{4}{3}d_d-\frac{1}{3}d_u.
\end{eqnarray}
We will assume the new physics contributions to $a_p$ and $d_n$ as
arising exclusively from the quark $u$, as in this case $f_j=t$.
Also, as usual, we will use $m_d\approx m_u\approx m_n/3\approx
m_p/3$. With these considerations, we can write the following
inequalities for the unknown parameters $Re(\Omega^2_{tu})$ and
$Im(\Omega^2_{tu})$:

\begin{eqnarray}
|Re(\Omega^2_{tu})|&<&(54\pi^2)\Bigg|\frac{\Delta
a^{Exp}_p}{x_tx_pg(x_t)}\Bigg|,\\
|Im(\Omega^2_{tu})|&<&(36\pi^2)\Big(\frac{m_H}{e}\Big)\Bigg|\frac{d^{Exp}_n}{x_tg(x_t)}\Bigg|.
\end{eqnarray}
Before analyzing more carefully the behavior of these parameters
with the Higgs mass, let us to estimate their size for a given
value of $m_H$. For instance, making $m_H=100$ GeV$/c^2$, a
straightforward evaluation shows that $|Re(\Omega^2_{tu})|\lesssim
6\times 10^{-3}$, whereas $|Im(\Omega^2_{tu})|\lesssim 10^{-7}$,
which means that the imaginary part of $\Omega_{tu}$ is more
suppressed than the real part. In the following, we will assume
that $Im(\Omega_{tu})\ll Re(\Omega_{tu})$ or that the Yukawa
sector is CP-conserving. After these considerations, we can write
\begin{equation}
\Omega^2_{tu}<(54\pi^2)\Bigg|\frac{\Delta
a^{Exp}_p}{x_tx_pg(x_t)}\Bigg|.
\end{equation}
In Figure~\ref{otu}, we show the behavior of this parameter as a
function of the Higgs mass. From this figure, it can be seen that
$\Omega^2_{tu}$ is practically insensitive to the variation of the
Higgs mass, as it go from $6\times 10^{-3}$ to $9\times 10^{-3}$
for $m_H$ ranging from $80$ GeV to $200$ GeV$/c^2$.

%%%%%%%%%%%%%%%%%%%%%%%%%%%%%%%%%%%%%%%%%%%%%%%%%%%%%%%%%%%%%%%%%%%%%%%%%%% FIGURE 2
%%%%%%%%%%%%%%%%%%%%%%%%%%%%%%%%%%%%%%%%%%%%%%%%%%%%%%%%%%%%%%%%%%%%%%%%%%%

Having obtained a bound for the $Htu$ vertex, we turn to comment
about the difficulties encountered in deriving a constraint on the
$Htc$ vertex. This vertex, in contrast with the $Htu$ one, has a
less direct connection with the electromagnetic properties of the
proton and the neutron, as it contributes to these quantities up
to the two-loop level. Due to the presence of the extra loop
factor $1/16\pi^2$, a factor proportional to $\alpha$, and the CKM
factor $|V_{cd}|^2$, the two-loop contribution of this coupling
to the $dd\gamma$ vertex is severally suppressed, which leads to
bounds for $|\Omega_{tc}|$ of about $10^{2}$. Definitively, it is
not possible to obtain an acceptable bound from low-energy data
for $|\Omega_{tc}|$, although from general considerations one
could expect that it is less restricted than $|\Omega_{tu}|$. In
this work, we will adopt a more conservative point of view by
assuming that $|\Omega_{tc}|$ obeys the same constraint than
$|\Omega_{tu}|$. Since the $m_u$ and $m_c$ masses will be ignored
through the calculations, it will not be possible to distinguish
among themselves the transitions $t\to u\tau \mu$ and $t\to
c\tau \mu$. Although our results would be more appropriated for
the former of these decays, in the following we will use the notation
$t\to u_i\tau \mu$. Notice that the bound for $\Omega_{tu_i}$ derived from the proton magnetic dipole moment is about one order of magnitude less stringent than that obtained from the $D^0-\overline{D^0}$ mixing.

It is worth comparing our results derived from low-energy data
with those that can be obtained from the Cheng-Sher ansatz~\cite{SHER1},
frequently used to make predictions in specific
flavor violating extended Yukawa sectors. According this ansatz,
the size of the $Htu_i$ vertex can be estimated by using the
relation $\lambda_{tu_i}\sqrt{m_tm_{u_i}}/v$. Assuming
$\lambda_{tu_{i}}$ of $O(1)$, one can estimate the values
$3\times 10^{-3}$ and $6\times10^{-2}$ for the $Htu$ and $Htc$
vertices, respectively. It is interesting to see that our bound for the $Htu_i$
vertex is of the same order of magnitude than that obtained from the
Cheng-Sher ansatz for $Htc$.

\subsubsection{Bounding $\Omega_{\tau \mu}$ from muon physics}
\begin{figure}
\centering
\includegraphics[width=4.0in]{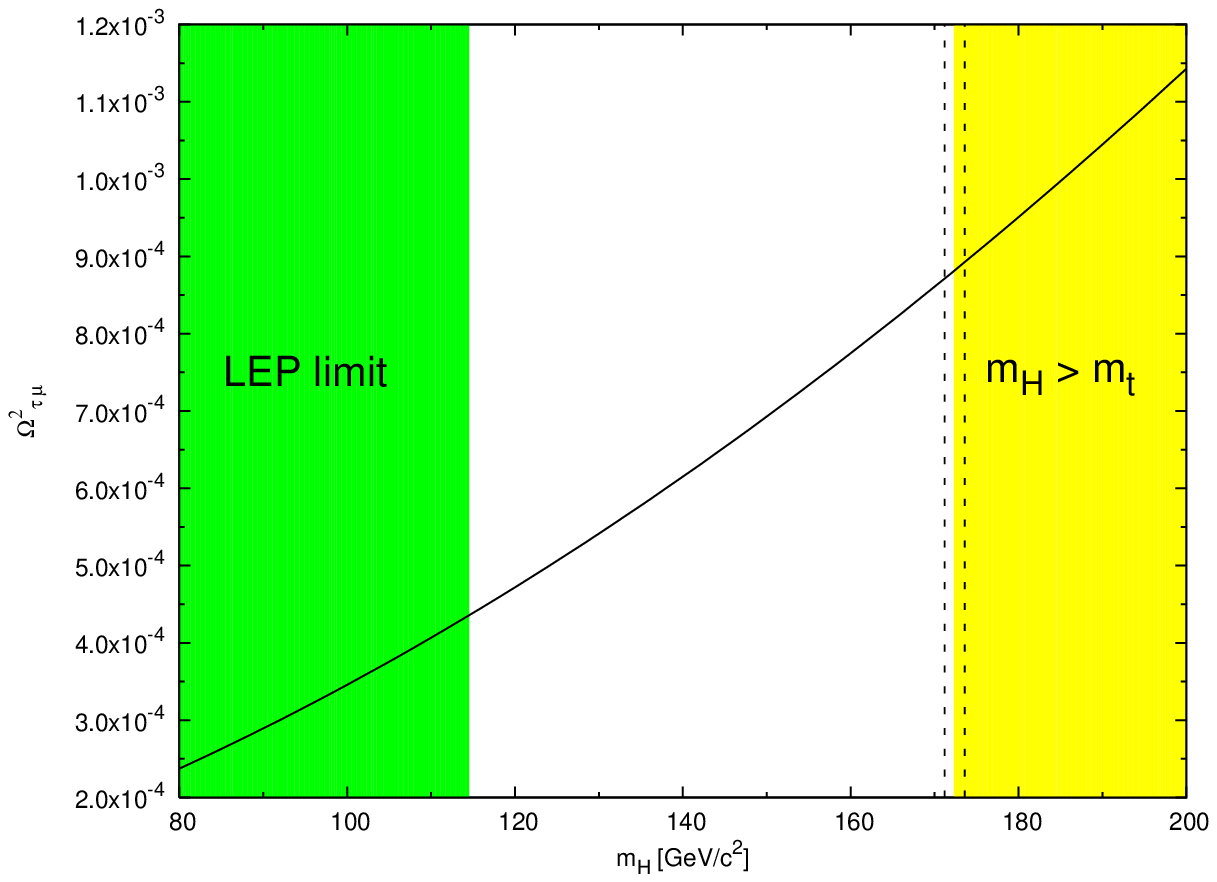}
\caption{\label{otm}Behavior of $\Omega^2_{\tau \mu}$ as a
function of the Higgs mass.}
\end{figure}

\begin{figure}
\centering
\includegraphics[width=2.0in]{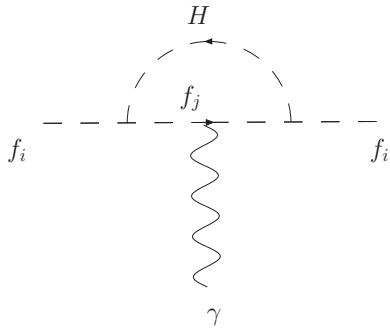}
\caption{\label{BT} Diagram contributing to the magnetic and
electric dipole moments of the $f_i$ fermion.}
\end{figure}

The anomalous magnetic moment of the muon is at present one of the
physical observables best measured. Here, we will use the
experimental uncertainty on this quantity to impose a bound on
$Re(\Omega^2_{\tau \mu})$. We will assume that the one-loop
contribution of the $H\tau \mu$ vertex is less than the experimental
uncertainty on this quantity, which is given by~\cite{PDG}:
\begin{equation}
|\Delta a^{Exp}_\mu|<5.4 \times 10^{-10}.
\end{equation}
As far as $Im(\Omega^2_{\tau \mu})$ is concerned, we can use the
existing experimental limit on the muon electric dipole moment,
which is given by~\cite{PDG}:
\begin{equation}
|d^{Exp}_\mu|<3.7\times 10^{-19} \ e.cm.
\end{equation}
From expressions given by Eqs.(\ref{md}) and (\ref{ed}),
one can write
\begin{eqnarray}
|Re(\Omega^2_{\tau \mu})|&<&(8\pi^2)\Bigg|\frac{\Delta
a^{Exp}_\mu}{x_\tau x_\mu g(x_\tau)}\Bigg|,\\
|Im(\Omega^2_{\tau
\mu})|&<&(16\pi^2)\Big(\frac{m_H}{e}\Big)\Bigg|\frac{d^{Exp}_\mu}{x_\tau
g(x_\tau)}\Bigg|.
\end{eqnarray}
The evaluation of these expressions for $m_H=100$ GeV$/c^2$ leads to
$|Re(\Omega^2_{\tau \mu})|\lesssim 3\times 10^{-4}$ and
$|Im(\Omega^2_{\tau \mu})|\lesssim 2$, showing that in this case
the imaginary part of $\Omega_{\tau \mu}$ is less constrained than
the real one. In order to simplify the analysis as much as
possible, we will assume that the leptonic Yukawa sector is also
CP-conserving. With these assumptions, we can write
\begin{equation}
\Omega^2_{\tau \mu}<(8\pi^2)\Bigg|\frac{\Delta a^{Exp}_\mu}{x_\tau
x_\mu g(x_\tau)}\Bigg|.
\end{equation}
In Figure~\ref{otm}, the behavior of $\Omega^2_{\tau \mu}$ as a
function of $m_H$ is shown. From this figure, it can be seen that
$\Omega^2_{\tau \mu}$ varies in approximately one order of
magnitude in the interval $80$ GeV$/c^2 < m_H< 200$ GeV$/c^2$.

%%%%%%%%%%%%%%%%%%%%%%%%%%%%%%%%%%%%%%%%%%%%%%%%%%%%%%%%%%%%%%%%%%%%%%%%%%% FIGURE 3
%%%%%%%%%%%%%%%%%%%%%%%%%%%%%%%%%%%%%%%%%%%%%%%%%%%%%%%%%%%%%%%%%%%%%%%%%%%
%%%%%%%%%%%%%%%%%%%%%%%%%%%%%%%%%%%%%%%%%%%%%%%%%%%%%%%%%%%%%%%%%%%%%%%%%%%

%%%%%%%%%%%%%%%%%%%%%%%%%%%%%%%%%%%%%%%%%%%%%%%%%%%%%%%%%%%%%%%%%%%%%%%%%%% FIGURE 4
%%%%%%%%%%%%%%%%%%%%%%%%%%%%%%%%%%%%%%%%%%%%%%%%%%%%%%%%%%%%%%%%%%%%%%%%%%%

%%%%%%%%%%%%%%%%%%%%%%%%%%%%%%%%%%%%%%%%%%%%%%%%%%%%%%%%%%%%%%%%%%%%%%%%%%%
\subsection{Bounds from lepton flavor violating transitions}
\begin{figure}
\centering
\includegraphics[width=3.0in]{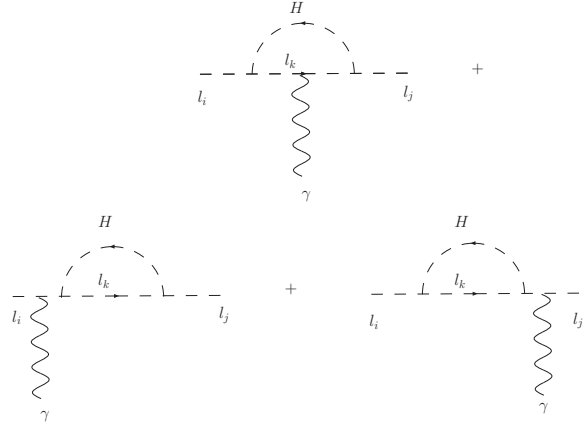}
\caption{\label{MD} Diagrams contributing to the $l_i \to
l_j\gamma$ decay.}
\end{figure}

In this subsection, we explore the possibility of deriving a bound
for the $H\tau \mu$ vertex from nondiagonal lepton
transitions~\cite{Pagliarone:2008af}. The best candidates are the $\tau \to \mu \bar{\mu} \mu $ and
$\tau \to \mu \gamma$ decays, to which there exist experimental
limits on their branching ratios. The current values reported by
the Particle Data Group~\cite{PDG} are Br$(\tau \to
\mu \bar{\mu} \mu)<3.2\times 10^{-8}$~\cite{Belle1} and Br$(\tau \to \mu
\gamma)<4.5\times 10^{-8}$~\cite{Belle2} at 90\% C.L. As we will see below, these
experimental limits are not enough to impose a competitive bound
on $|\Omega_{\tau \mu}|$. On the other hand, the one-loop flavor
violating decay $\mu \to e\gamma$, to which the $\tau$ lepton can
contribute virtually inside a loop (see Figure~\ref{MD}), offers
by far the better option to bound flavor violating transitions, as
it is more strongly restricted by the experiment. The
corresponding experimental limit reported by the Particle Data
Group~\cite{PDG} is: $Br(\mu \to e\gamma)<1.2\times 10^{-11}$ at
90\% C.L. The decay $\mu \to e\bar{e}e$ is also strongly
constrained by the experiment, but it can not be used to
constraint the $H\tau \mu$ vertex at the tree level. As far as the
$H\tau \mu$ contribution to the one-loop $\mu \to e\gamma$ decay,
it is important to note that one needs to introduce also the
$H\tau e$ coupling. Thus, it is not possible to derive directly a
bound for the $\Omega_{\tau \mu}$ parameter in an isolated way,
but only through the $\Omega_{\tau \mu}\Omega_{\tau e}$ product.
As we will see below, a very restrictive bound on $\Omega_{\tau
\mu}\Omega_{\tau e}$ can be derived, but due to the presence of
the $\Omega_{\tau e}$ parameter, additional assumptions are needed
in order to make predictions.

After these general considerations, we proceed to examine each of
the processes above mentioned. We begin with the tree level
three-body decay $\tau \to \mu \bar{\mu}\mu$, which occurs
through a diagram similar to the one shown in Figure~\ref{TD}.
Assuming a $H\mu \mu$ vertex as the one appearing in the SM, one
obtains:
\begin{equation}
|\Omega_{\tau
\mu}|^2<\Big(\frac{256\pi^2s^2_W}{\alpha}\Big)\Big(\frac{m_W}{m_\mu}\Big)^2\Big(\frac{\Gamma_\tau}{m_\tau}\Big)
\frac{Br^{Exp}(\tau \to \mu \bar{\mu} \mu)}{f(y,z)},
\end{equation}
where $\Gamma_\tau$ is total $\tau$ decay width, whereas the
$f(y,z)$ function is given by Eq.~(\ref{ff}), but in this case
$y=m^2_H/m^2_\tau$ and $z=\Gamma^2_H/m^2_\tau$. A straightforward
evaluation for $m_H=100$ GeV$/c^2$ leads to
\begin{equation}
|\Omega_{\tau \mu}|^2<0.23,
\end{equation}
which is much less restrictive than that obtained from the muon anomalous
magnetic dipole moment. It is worth mentioning that an identical
bound is obtained for the $\Omega_{\tau e}$ parameter using the
experimental constraint on the $\tau \to e\bar{\mu}\mu$
transition, whose limit on its branching ratio is~\cite{Babar1}
Br$(\tau \to e \bar{\mu} \mu)<3.7\times 10^{-8}$ at 90\% C.L.

We now turn to analyze the one-loop decay $l_i\to l_j\gamma$,
with $l_i$ stands for a charged lepton. The contribution of the
flavor violating $Hl_il_k$ and $Hl_k l_j$ vertices to the $l_i \to
l_j\gamma$ decay is given through the diagrams shown in
Figure~\ref{MD}. The corresponding amplitude can be written as
\begin{equation}
{\cal M}(l_i \to l_j\gamma)=\bar{u}(p_2)\Gamma_\mu u(p_1)
\epsilon^{*\mu}(q,\lambda),
\end{equation}
where the vertex function is given by
\begin{equation}
\Gamma_\mu=e\Omega_{ki}\Omega_{kj}\int
\frac{d^Dk}{(2\pi^2)^D}\sum^3_{a=1}\frac{T^a_\mu}{\Delta_a},
\end{equation}
with
\begin{eqnarray}
T^1_\mu&=&(\pFMSlash{k}+\pFMSlash{p_2}+m_k)\gamma_\mu
(\pFMSlash{k}+\pFMSlash{p_1}+m_k),\\
T^2_\mu&=&-\frac{1}{m^2_i}(\pFMSlash{k}+\pFMSlash{p_2}+m_k)(\pFMSlash{p_2}+m_i)\gamma_\mu,
\\
T^3_\mu&=&\frac{1}{m^2_i}\gamma_\mu\pFMSlash{p_1}(\pFMSlash{k}+\pFMSlash{p_1}+m_k),
\end{eqnarray}
\begin{eqnarray}
\Delta_1&=&[k^2-m^2_H][(k+p_1)^2-m^2_k][(k+p_2)^2-m^2_k],\\
\Delta_2&=&[k^2-m^2_H][(k+p_2)^2-m^2_k],\\
\Delta_3&=&[k^2-m^2_H][(k+p_1)^2-m^2_k].
\end{eqnarray}
In deriving the above expressions, a CP conserving Yukawa sector
has been assumed. Also, the $m_j$ mass has been neglected. There
is no contribution to the charge, as it should be, but only to the
transition magnetic dipole moment. This contribution is given by
\begin{equation}
\Gamma_\mu=-\frac{e\Omega_{ki}\Omega_{kj}}{32\pi^2 m_\mu}{\cal
A}\sigma_{\mu \nu} q^\nu,
\end{equation}
where the loop function ${\cal A}$ is given by
\begin{eqnarray}
{\cal A}&=&\frac{1}{2}+m^2_k
\Big(1+\frac{m_i}{m_k}\Big)C_0+\Big(\frac{m_H}{m_i}\Big)^2\Bigg(1-\Big(\frac{m_i}{m_H}\Big)\Big(\frac{m_k}{m_H}\Big)\nonumber\\
&&-\Big(\frac{m_k}{m_H}\Big)^2\Bigg)\Big(B_0(1)-B_0(2)\Big).
\end{eqnarray}
Here, $B_0(i)$ and $C_0$ are Passarino-Veltman scalar
functions~\cite{PaVe}, which are given by
$C_0=C_0(m^2_i,0,0,m^2_H,m^2_k,m^2_k)$,
$B_0(1)=B_0(0,m^2_H,m^2_k)$, and $B_0(2)=B_0(m^2_i,m^2_H,m^2_k)$.
After squaring the amplitude, one obtains for the branching ratio:
\begin{equation}
Br(l_i \to l_j\gamma)=\frac{\alpha \Omega^2_{ki}\Omega^2_{k
j}}{8(4\pi)^4}\Bigg(\frac{m_i}{\Gamma_i}\Bigg)|{\cal A}|^2,
\end{equation}
where in this expression $\Gamma_i$ stands for the $l_i$ lepton
decay width. This branching ratio must be less than the
experimental limit, so this condition translates into a constraint
for the $\Omega_{ki}\Omega_{kj}$ product,
\begin{equation}
|\Omega_{ik}\Omega_{k
j}|<\sqrt{\frac{8(4\pi)^4}{\alpha}\Bigg(\frac{\Gamma_i}{m_i}\Bigg)\Bigg(\frac{Br_{Exp}(l_i
\to l_j\gamma)}{|{\cal A}|^2}\Bigg)}.
\end{equation}
To constraint $\Omega_{\tau \mu}$, we take in the above expression
$l_i=l_k=\tau$ and $l_j=\mu$, with the $H\tau \tau$ coupling as in
the SM. Once introduced these changes in the above general result,
one obtains
\begin{equation}
|\Omega_{\tau \mu}|<\frac{64\pi
s_W}{\alpha}\Big(\frac{m_W}{m_\tau}\Big)\sqrt{\frac{\pi}{2}\Bigg(\frac{\Gamma_\tau}{m_\tau}\Bigg)
\Bigg(\frac{Br_{Exp}(\tau \to \mu \gamma)}{|{\cal A}|^2}\Bigg)},
\end{equation}
which ($m_H=100$ GeV$/c^2$) leads to the following constraint:
\begin{equation}
|\Omega_{\tau \mu}|<10^{-1},
\end{equation}
which is about one order of magnitude less stringent than that
obtained from the muon magnetic dipole moment.

\begin{figure}
\centering
\includegraphics[width=4.0in]{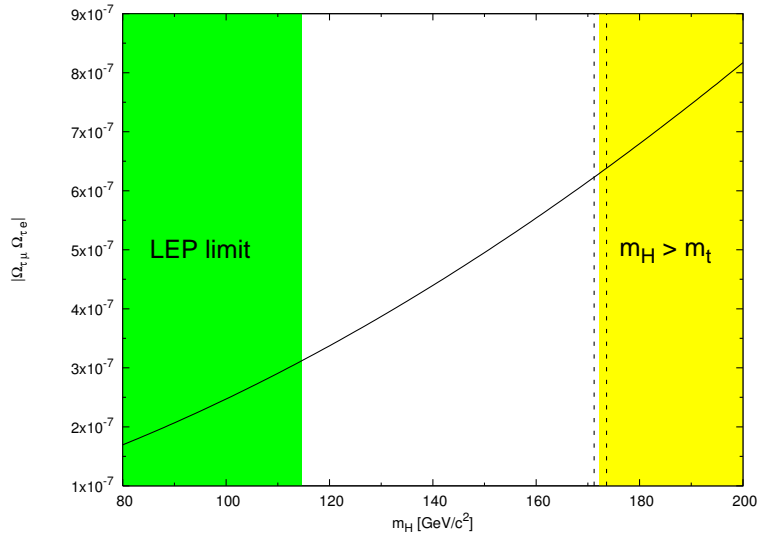}
\caption{\label{FIG1} Allowed values for the product
$|\Omega_{\tau \mu}\Omega_{\mu e}|$ as a function of the Higgs
mass.}
%the two shaded areas represent the LEP lower limit on the SM
%Higgs mass~\cite{LEPBound} and the region in which the decay $t
%\rightarrow u_i H$ is kinematically
%forbidden~\cite{tevatrontopmass}.}
\end{figure}

Although it is not possible to obtain an acceptable constraint for
the $\Omega_{\tau \mu}$ parameter, it is interesting to analyze
the constraint induced by the experimental limit on the $\mu \to
e\gamma$ decay on the $|\Omega_{\tau \mu}\Omega_{\tau e}|$
product. In Figure~\ref{FIG1}, its behavior as a function of $m_H$
is displayed. For clarity, this variation is shown for $m_H$
ranging from $80$ to $200$ GeV$/c^2$. It can be appreciated from
this figure that $|\Omega_{\tau \mu}\Omega_{\tau e}|$ varies in
this interval from approximately $1.5\times 10^{-7}$ to $8\times
10^{-7}$. Although very stringent, this constraint cannot be used
to predict the $H\tau \mu$ vertex without making additional
assumptions concerning the $\Omega_{\tau e}$ parameter. However, if we adopt an approach similar to the one used for the $\Omega_{tu}\Omega_{tc}$ product, a very restrictive branching ratio for the $t\to u_i\tau l_i$  decay is obtained, as the corresponding bound for $\Omega_{\tau l_i}$ ($l_i=e,\mu$) is more stringent than that obtained from the muon anomalous magnetic moment by more than two orders of magnitude. Although we prefer to make predictions for the $t\to u_i\tau \mu$ transition using the constraint for $\Omega_{\tau \mu}$  obtained from the muon magnetic moment, as it does not require of introducing extra assumptions, the scenario for a very constrained $t\to u_i\tau l_i$ decay will be also discussed. Indeed, this is a good approach to predict the branching ratio of the $t\to u_i\tau e$ transition but not for the  $t\to u_i\tau \mu$, as it is natural to expect that $\Omega_{\tau \mu}>\Omega_{\tau e}$. The possibility of considering some type of relations between the $\Omega_{\tau \mu}$ and $\Omega_{\tau e}$ parameters will be not considered in this work because, as it is emphasized in the introduction, our study will be supported only in the available experimental data. On the other hand, it should
be commented that a similar analysis to the presented here was
carried out some years ago~\cite{SHER2} to constraint the
$\lambda_{ij}$ parameters appearing in the Cheng-Sher ansatz
within the context of multi-Higgs extensions of the SM. We have
verified that our results are in perfect agreement with those
given in this reference.

To finish this section, it is interesting to compare our bound for
$\Omega_{\tau \mu}$ with than that can be derived from the
Cheng-Sher ansatz. In this case, $\Omega_{\tau \mu}<\sqrt{m_\tau
m_\mu}/v \approx 1.7\times 10^{-3}$, which shows that our bound
obtained from the muon anomalous magnetic dipole moment is less
restrictive by about one order of magnitude.

%%%%%%%%%%%%%%%%%%%%%%%%%%%%%%%%%%%%%%%%%%%%%%%%%%%%%%%%%%%%%%%%%%%%%%%%%%% FIGURE 5
%%%%%%%%%%%%%%%%%%%%%%%%%%%%%%%%%%%%%%%%%%%%%%%%%%%%%%%%%%%%%%%%%%%%%%%%%%%

%%%%%%%%%%%%%%%%%%%%%%%%%%%%%%%%%%%%%%%%%%%%%%%%%%%%%%%%%%%%%%%%%%%%%%%%%%% FIGURE 6
%%%%%%%%%%%%%%%%%%%%%%%%%%%%%%%%%%%%%%%%%%%%%%%%%%%%%%%%%%%%%%%%%%%%%%%%%%%

%%%%%%%%%%%%%%%%%%%%%%%%%%%%%%%%%%%%%%%%%%%%%%%%%%%%%%%%%%%%%%%%%%%%%%%%%%%
\section{Results and discussion}
\label{r}In this section, we discuss our results. As commented at the end of the previous section, we will discuss the branching ratio of the $t\to u_i\tau \mu$ decay using the constraint for the $H\tau \mu$ vertex obtained from the muon anomalous magnetic moment, and also for the branching ratio of the $t\to u_il_i$ transition using a bound for the $H\tau l_i$ coupling derived from the experimental limit on the $\mu \to e\gamma$ decay. In the latter case, our result is a good prediction for the $t\to u_i\tau e$ decay, but not for the $t\to u_i \tau \mu$ one, as it is natural to assume that the $\Omega_{\tau \mu}>\Omega_{\tau e}$ relation is satisfied in the fundamental theory.

\subsection{Decay $t\to u_i\tau \mu$}
\begin{figure}
\centering
\includegraphics[width=4.5in]{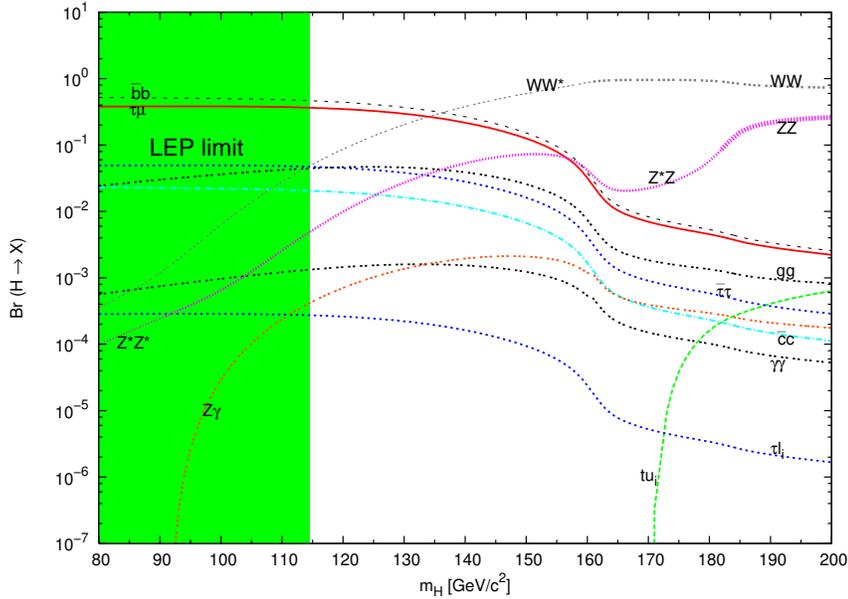}
\caption{\label{BRH}Branching ratios for the Higgs boson decay into the  $\bar bb, \bar\tau\tau,\bar cc,gg,Z\gamma,WW^\ast,WW,Z^\ast Z^\ast,Z^\ast Z,ZZ,\tau \mu,\tau l_i$, and $tu_i$ modes as a function of $m_H$.}
\end{figure}

\begin{figure}
\centering
\includegraphics[width=3.5in]{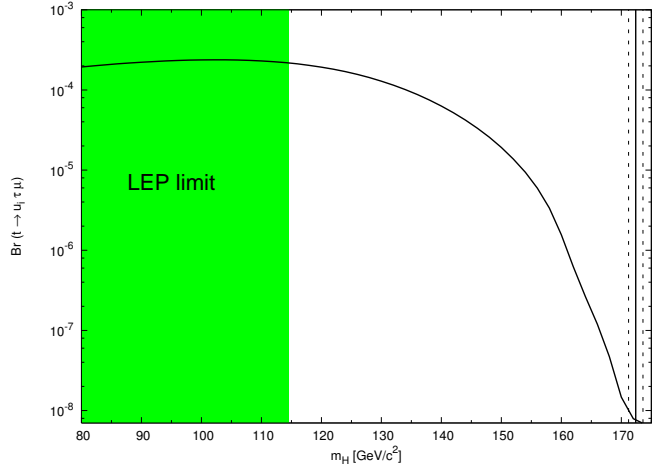}
\caption{\label{BR}Branching ratio for the $t\to u_i\tau \mu$
decay as a function of the Higgs mass.}
\end{figure}

When the $t\to u_i\tau \mu$ transition is mediated by the $H$ scalar resonance,
the highest value that can reach the corresponding branching ratio
can be expressed simply as
\begin{equation}
Br(t\to u_i\tau \mu)=Br(t\to u_iH)Br(H\to \tau \mu),
\end{equation}
where
\begin{eqnarray}
Br(t\to
u_iH)&=&\frac{\Omega^2_{tu}}{32\pi}\Bigg(\frac{m_t}{\Gamma_t}\Bigg)\left(1-\Big(\frac{m_H}{m_t}\Big)^2\right)^2\\
Br(H\to \tau \mu)&=&\frac{\Omega^2_{\tau
\mu}}{4\pi}\Bigg(\frac{m_H}{\Gamma_H}\Bigg)\left(1-\Big(\frac{m_\tau}{m_H}\Big)^2\right)^2
\end{eqnarray}
In writing these expressions we have ignored the $u_i$ and $\mu$
masses. These branching ratios are of interest by themselves, so
we will evaluate them for an allowed value of the Higgs mass that
maximize their values.
%%%
As it can be seen from Figure~\ref{BR}, the branching ratio for
the decay $t \to u_i\tau \mu$ reach its highest value around
$m_H=\,105$ GeV$/c^2$, that is below the region allowed for the present
analysis ($114.4$ GeV$/c^2$ $ <\,m_H\,<\, 171.3$ ($\pm 1.1$)
GeV$/c^2$).
%%%
Specifically, to estimate these branching ratios, we will take
$m_H=115$ GeV$/c^2$. In addition, we will use for the total top
decay width the approximation $\Gamma_t\approx \Gamma(t \to
bW)=1.55$ GeV$/c^2$. Using these values and the constraint
$\Omega^2_{tu}<1.03\times 10^{-3}$ derived from Eq.~(\ref{CDD}), one obtains $Br(t\to Hu_i)\sim
3.7\times 10^{-4}$. As far as $Br(H\to \tau \mu)$ is concerned,
we will use the constraint $\Omega^2_{\tau \mu}<2\times 10^{-4}$ and a total Higgs decay width given by: $\Gamma_H= \Gamma (H\to \mathrm{all\,\, SM\,\, decay\,\, modes})+\Gamma(H\to \tau\mu)+\Gamma(H\to \tau l_i)$~\footnote{The total decay width for all SM-Higgs decay modes was evaluated using the current version of the HDECAY program~\cite{HDECAY}.}. In  Fig.~\ref{BRH}, the behavior of the branching ratios for the  $H\to \tau \mu$ and $H\to \tau l_i$ decays are shown as a function of the Higgs mass. In order to appreciate their relative importance, the branching ratios associated with the most representative SM Higgs decays are shown too. It can be appreciated from this figure that $Br(H\to \tau \mu)$ ranges from $0.37$ to $0.04$ in the interval $115$ GeV $<m_H< 2m_W$. In this way, the branching ratio for the $t \to u_i\tau \mu$ decay can reach the value $1.37 \times 10^{-4}$, at the best.
%%%%%%%%%%%%%%%%%%%%%%%%%%%%%%%%%%%%%%%%%%%%%%%%%%%%%%%%%%%%%%%%%%%%%%%%%%% FIGURE 7
%%%%%%%%%%%%%%%%%%%%%%%%%%%%%%%%%%%%%%%%%%%%%%%%%%%%%%%%%%%%%%%%%%%%%%%%%%% FIGURE 7
%%%%%%%%%%%%%%%%%%%%%%%%%%%%%%%%%%%%%%%%%%%%%%%%%%%%%%%%%%%%%%%%%%%%%%%%%%%
%%%%%%%%%%%%%%%%%%%%%%%%%%%%%%%%%%%%%%%%%%%%%%%%%%%%%%%%%%%%%%%%%%%%%%%%%%%

Before discussing our result for the branching ratio of the $t\to
u_i\tau \mu$ decay, it is worth comparing our prediction for the
two-body $t \to u_iH$ and $H\to \tau \mu$ decays with those found
in other contexts. We will focus on the rare top decay $t\to cH$,
as the $t\to uH$ mode is expected to be much more suppressed. In
the SM, this decay is quite suppressed to be detected, as it has a
branching ratio ranging from $10^{-13}$ to $10^{-14}$ for
$m_Z<m_H<2m_W$~\cite{Mele}. Although very suppressed in the SM,
this decay can reach sizable branching ratios in many well-motivated
SM extensions. For instance, it can reach branching
ratios ranging from $10^{-3}$ to $10^{-7}$ in the general
two-Higgs doublet models~\cite{TcHTHDM} or in the minimal
supersymmetric standard model~\cite{TcHMSSM}. As to the lepton
flavor violating Higgs decays is concerned, they are absent at any
order of perturbation theory in the SM, but they can be induced
with sizable branching ratios in many of its extensions. For
instance, the decay $H\to \tau \mu$ can reach branching ratios as
large as $10^{-3}-10^{-2}$ in the THDM-III~\cite{HLFVTHDM}, in the
MSSM~\cite{HLFVMSSM}, or in a specific SUSY-$SU(5)$ scenario~\cite{T3}.
Motivated by the strong experimental evidence of
nonzero masses for the light neutrinos~\cite{NO}, important
attention has received the possible violation of the lepton flavor
mediated by the Higgs boson in the context of seesaw models.
Although these decays are very suppressed within the SM-seesaw
model~\cite{SM-SEESAW, HERRERO}, with branching ratios of order of
$10^{-56}$, they can dramatically be enhanced in other models, as
the MSSM-seesaw model, in which can reach values as large as
$10^{-5}$~\cite{HERRERO}.

We now turn to discuss the branching ratio for the $t \to u_i\tau
\mu$ decay. As it can be appreciated from Figure~\ref{BR}, this
branching ratio reach its highest value of approximately $10^{-4}$
for $m_H=105$ GeV$/c^2$, but we are interested in examining its
behavior for values of $m_H$ above the LEP limit. It can be seen
from this figure that for $115<m_H<2m_W$, $Br(t\to u_i\tau \mu)$
varies from approximately $10^{-4}$ to $10^{-5}$. However, the
branching ratio decreases strongly for $m_H>2m_W$, which means
that this decay is only interesting when mediated by a relatively
light Higgs boson. The importance of this branching ratio can be
best appreciated if compared with those existing in the literature
for other top quark three-body decays. In the SM, the three-body
decays of the top quark are in general very suppressed. For
instance, the flavor violating decays $t\to cWW$~\cite{Elizabeth},
$t\to u_i\bar{u}_ju_j$~\cite{T1,Mariana1}, and $t\to
cgg$~\cite{Mariana1} have branching ratios of order of $10^{-12}$,
$3.4\times 10^{-12}$, and $10^{-9}$, respectively. Interestingly,
it was found~\cite{T1,Mariana1} that the latter two decays have
branching ratios as large as or larger than those associated with
the two-body decay $t\to u_ig$~\cite{TopLoop}. However, the
decays $t \to cV_iV_j$ ($V_i=W,Z,\gamma,g$) can have sizable
branching ratios of order of $10^{-4}$ to $10^{-6}$ in the
THDM~\cite{THDMtvv} and the MSSM~\cite{MSSMtvv}. On the other
hand, branching ratios of order of $10^{-8}-10^{-5}$ has been
determined in the context of the THDM-III~\cite{Iltan, Mariana2}
and the MSSM~\cite{Mariana2} for the quark flavor violating, but
lepton flavor conserving, $t\to cl^+l^-$ decay. The double flavor
violating $t\to u_i\tau \mu$ decay has thus an important branching
ratio within this category of rare top quark decays.

%%%%%%%%%%%%%%%%%%%%%%%%%%%%%%%%%%%%%%%%%%%%%%%%%%%%%%%%%%%%%%%%%%%%%%%%%%% FIGURE 8
%%%%%%%%%%%%%%%%%%%%%%%%%%%%%%%%%%%%%%%%%%%%%%%%%%%%%%%%%%%%%%%%%%%%%%%%%%% FIGURE 8
%%%%%%%%%%%%%%%%%%%%%%%%%%%%%%%%%%%%%%%%%%%%%%%%%%%%%%%%%%%%%%%%%%%%%%%%%%%
%%%%%%%%%%%%%%%%%%%%%%%%%%%%%%%%%%%%%%%%%%%%%%%%%%%%%%%%%%%%%%%%%%%%%%%%%%%

\subsection{Decay $t\to u_i\tau l_i$}
\begin{figure}
\centering
\includegraphics[width=3.5in]{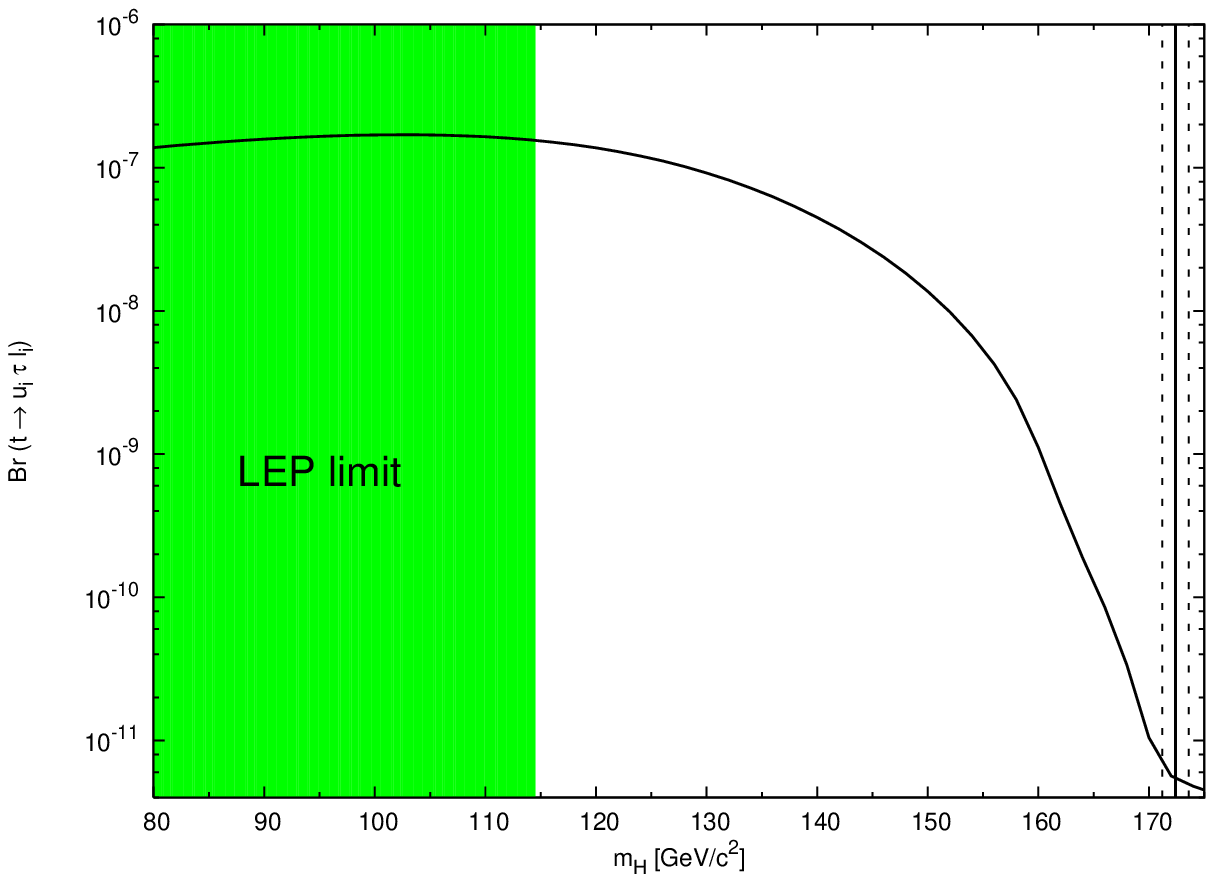}
\caption{\label{BRli}Branching ratio for the $t\to ui\tau l_i$
decay as a function of the Higgs mass.}
\end{figure}

This subsection is devoted to discuss the branching ratio for the $t\to u_i\tau l_i$ decay using the constraint for the $H\tau l_i$ vertex derived from the experimental limit on the $\mu \to e\gamma$ transition. As mentioned at the end of the previous section, our result is a good prediction for the $t\to u_i\tau e$ decay, but not for the $t\to u_i\tau \mu$ one, as it is expected that the $H\tau e$ vertex is much more suppressed than the $H\tau \mu$ one. The behavior of $Br(H\to \tau l_i)$ as a function of the Higgs mass is shown in Fig.~\ref{BRH}. From this figure, it can be appreciated that this branching ratio ranges from $2.7\times 10^{-4}$ to $3.1\times 10^{-5}$ for a Higgs mass in the interval $115$ GeV $<m_H<2m_W$. Using the bound for $Br(t\to u_iH)$ derived from the $D^0-\bar{D}^0$ mixing, it is found that $Br(t\to u_i\tau l_i)=1.05\times 10^{-7}$, at the best. Its behavior as a function of the Higgs mass is shown in Fig.~\ref{BRli}. Strictly speaking, this result can be considered as a stringent bound for the $t\to u\tau e$ decay, in whose derivation the conservative point of view of assuming $\Omega_{tc}=\Omega_{tu}$ and $\Omega_{\tau \mu}=\Omega_{\tau e}$ was adopted. This branching ratio is too much small to be at the reach of the LHC sensitivity.

%%%%%%%%%%%%%%%%%%%%%%%%%%%%%%%%%%%%%%%%%%%%%%%%%%%%%%%%%%%%%%%%%%%%%%%%%%% FIGURE 9
%%%%%%%%%%%%%%%%%%%%%%%%%%%%%%%%%%%%%%%%%%%%%%%%%%%%%%%%%%%%%%%%%%%%%%%%%%% FIGURE 9
%\begin{figure}
%\centering
%\includegraphics[width=3.5in]{BRhtauli.eps}
%\caption{\label{BRH2}Branching ratio for the $H\to \tau l_i$
%decay as a function of the Higgs mass. The $H\to WW^*$ decay mode has been included.}
%\end{figure}
%%%%%%%%%%%%%%%%%%%%%%%%%%%%%%%%%%%%%%%%%%%%%%%%%%%%%%%%%%%%%%%%%%%%%%%%%%%
%%%%%%%%%%%%%%%%%%%%%%%%%%%%%%%%%%%%%%%%%%%%%%%%%%%%%%%%%%%%%%%%%%%%%%%%%%%

%%%%%%%%%%%%%%%%%%%%%%%%%%%%%%%%%%%%%%%%%%%%%%%%%%%%%%%%%%%%%%%%%%%%%%%%%%% FIGURE 10
%%%%%%%%%%%%%%%%%%%%%%%%%%%%%%%%%%%%%%%%%%%%%%%%%%%%%%%%%%%%%%%%%%%%%%%%%%% FIGURE 10
%%%%%%%%%%%%%%%%%%%%%%%%%%%%%%%%%%%%%%%%%%%%%%%%%%%%%%%%%%%%%%%%%%%%%%%%%%%
%%%%%%%%%%%%%%%%%%%%%%%%%%%%%%%%%%%%%%%%%%%%%%%%%%%%%%%%%%%%%%%%%%%%%%%%%%%

%%%%%%%%%%%%%%%%%%%%%%%%%%%%%%%%%%%%%%%%%%%%%%%%%%%%%%%%%%%%%%%%%%%%%%%%%%%%%%%%%%%%
\subsection{Experimental perspectives}
%%%%%%%%%%%%%%%%%%%%%%%%%%%%%%%%%%%%%%%%%%%%%%%%%%%%%%%%%%%%%%%%%%%%%%%%%%%%%%%%%%%%

The Large Hadron Collider will be an ideal machine for investigating the characteristic of
the heaviest quark and its role in the SM.
%%%%%%%%
With a NLO production cross-section of about $830$ $pb$, two top-pairs
per second are expected to be produced. This means that we should be able
to collect more than $10$ millions of $t \bar t$ per year at low
luminosity conditions of around $2\times 10^{33}\,cm^2\,s^{-1}$.
%%%%%%%%
The $t \bar t$ production at the LHC is predicted to take place via two distict hard processes, the
$gg \rightarrow t \bar t$, wich will contribute for about $90$\% to the total production cross
section and $g \bar q  \rightarrow t \bar t$ process which contributes for the remaining $10$\%.
%%%%%%%%

%%%%%%%%
Within the SM, a top, with a mass above $Wb$ threshold, is predicted
to have a decay width dominated by the two-body process: $t \rightarrow W b\,\,$
($Br\simeq 0.998$)~\cite{PDG}.
%%%%%%%%
A $t \bar t$ pair will then decay via one of the following channels,
classified according to the final states:
%%%%%%%%
$i)$ dilepton, where both $W$ decays are leptonic ($\ell_{W}$),
with 2 jets arising from the two $b$-quark hadronization and
missing transverse energy ($\not\!\!E_{\rm T}$)~\cite{MET} coming
from the undetected neutrinos:
$Br(e,e)=Br(\mu,\mu)=Br(\tau,\tau) \simeq 1/81$, $Br(e,\mu)=Br(e,\tau)=Br(\mu,\tau) \simeq 2/81$;
%%%%%%%%%%%%%%%%%%%%%%%%%%%%%%%%%%%%%%%%%%%%%%%%%%%%%%%%%%%%%%%%%%%%%%%%%%%%%%%%%%%%
$ii)$ lepton$\,+\,$jets, where one $W$ decays leptonically and the other one
into quarks, with 4 jets and $\not\!\!E_{\rm T}$:
%The SM branching ratios in these cases are:
$Br(e+jets)=Br(\mu+jets)=Br(\tau+jets) \simeq 12/81$;
%%%%%%%%%%%%%%%%%%%%%%%%%%%%%%%%%%%%%%%%%%%%%%%%%%%%%%%%%%%%%%%%%%%%%%%%%%%%%%%%%%%%
$iii)$ all jets, where both the $W$'s decay into quarks with $6$
jets and no associated $\not\!\!E_{\rm T}$;
the corresponding branching fraction is $Br(jets) \simeq 36/81$.
%%%%%%%%

%%\noindent
If $\mathcal{B}$ represents the branching ratio for the Higgs mediated double flavor violating top
quark decay: $\mathcal{B} \equiv$ Br$(t\to u_i\tau \mu)$ and if we assume no other
significantly accessible decay channels, either than the SM and the DFV, we will have:
Br$(t \rightarrow W b)= 1 - \mathcal{B}$.
%%%%%%%%%%%%%%%%%%%%%%%%%%%%%%%%%%%%%%%%%%%%%%%%%%%%%%%%%%%%%%%%%%%%%%%%%%%%%%%%%%%%%%%%%%%%%%%%%%%
%%%
A $t \bar{t}$ pair can then decays as follows:
a) purely SM decays (SM-SM), where both the top quarks decay into $Wb$ (Br$(t \bar t)|_{SM}=(1- \mathcal{B})^2$);
b) mixed SM-DFV decays, where one top will decay into $Wb$ and the other into $u_i\, \mu \, \tau$
(Br$(t \bar t)|_{SM-DFV}=$ $2{\mathcal{B}}(1-{\mathcal{B}})$);
c) purely DFV decays (DFV-DFV), in this case both the top quarks will decay into $u_i\, \mu \, \tau$
($Br(t\bar t)|_{DFV}= \mathcal{B}^2$).
%%%
As a purely DFV $t \bar t$ decay is strongly suppressed
($B\simeq$ $10^{-6}-10^{-8}$), we will focus only on SM-DFV top quark decays.
%%

%%%%   TAU ID
%%%%%%%%%%%%%%%%%%%%%%%%%%%%%%%%%%%%%%%%%%%%%%%%%%%%%%%%%%%%%%%%%%%%%%%%%%%%%%%%%%%%%%%%%%%%%%%%%%%%
%%\noindent
The $\tau$ leptons, produced by the DFV top quark, will decay either in
$e\overline{\nu}_{e}\nu_{\tau}$ or $\mu\overline{\nu}_{\mu}\nu_{\tau}$ ($\tau_{\ell}$)
or to hadrons ($\tau_{h}$).
%%%%
The branching ratios for such decay processes are:
$\Gamma^{\tau}(\mu^{-} \overline{\nu}_{\mu} \nu_{\tau})/\Gamma^{\tau}_{Total}= 0.1736 \pm 0.005$,
$\Gamma^{\tau}(e^{-} \overline{\nu}_{e} \nu_{\tau})/\Gamma^{\tau}_{Total}= 0.1784 \pm 0.005$
and $\Gamma^{\tau}(\tau \to jet)/\Gamma^{\tau}_{Total} \simeq 0.653$~\cite{PDG}.
%%%%%%%%%%%%%%%%%%%%%%%%%%%%%%%%%%%%%%%%%%%%%%%%%%%%%%%%%%%%%%%%%%%%%%%%%%%%%%%%%%%%%%%%%%%%%%%%%%%
In $\simeq 77$ \% of hadronic $\tau$ decays, only one charged
track is produced: $\tau \rightarrow \nu_\tau + 3\pi^\pm +n \pi^0$ ($1$-prong decay); in the other
$\simeq 23$ \% of cases we will have $3$ charged tracks: $\tau \rightarrow \nu_\tau + 3\pi^\pm +n \pi^0$
($3$-prong decay).
%%%%%%%%%%%%%%%%%%%%%%%%%%%%%%%%%%%%%%%%%%%%%%%%%%%%%%%%%%%%%%%%%%%%%%%%%%%%%%%%%%%%%%%%%%%%%%%%%%%
A $\tau$ lepton, decaying hadronically, will generate a small jet ($\tau_h$) with hadrons
and neutrinos amongst the decay products. The presence of such a particles makes difficult
to reconstruct and identify efficiently the $\tau$-jets.
The background misidentified as a $\tau_h$ is mainly QCD multi jet events, but also electrons
that shower late or with strong bremsstrahlung, or muons interacting in the calorimeters.
When the momentum of the $\tau$ is large, compared to the mass, a very collimated jet will
be produced. Then, a $\tau$-jet can be identified through the presence of a well collimated
calorimeter cluster with a small number of associated charged traks ($1$ to $3$).
%%%%%%%%%%%%%%%%%%%%%%%%%%%%%%%%%%%%%%%%%%%%%%%%%%%%%%%%%%%%%%%%%%%%%%%%%%%%%%%%%%%%%%%%%%%%%%%
At LHC center of mass energy,
for a transverse momentum $P_T>$ $50$ GeV$/c$,
90\% of the energy is contained in a cone of radius $R=\sqrt{ \Delta \eta^2 + \Delta \Phi^2}=$ $0.2$.
Hadronic $\tau$ decays have low charged track multiplicity and a relevant
fraction of electromagnetic energy deposition in the calorimeters due to photons coming from
the decay of neutral pions.
%%%%%%%%%%%%%%%%%%%%%%%%%%%%%%%%%%%%%%%%%%%%%%%%%%%%%%%%%%%%%%%%%%%%%%%%%%%%%%%%%%%%%%%%%%%%%%%%%%%

%% Analysis Strategy
%%%%%%%%%%%%%%%%%%%%%%%%%%%%%%%%%%%%%%%%%%%%%%%%%%%%%%%%%%%%%%%%%%%%%%%%%%%%%%%%%%%%%%%%%%%%%%%%%%%
%%% EXPERIMENTAL SIGNATURE
%%%%%%%%%%%%%%%%%%%%%%%%%%%%%%%%%%%%%%%%%%%%%%%%%%%%%%%%%%%%%%%%%%%%%%%%%%%%%%%%%%%%%%%%%%%%%%%%%%%%
As in the channel $t \bar t \to u_i\tau \mu W b$ there is no charge correlation between the $\mu$
and the $W$, coming from the top decay, if the $W$ decays leptonically ($W \rightarrow \ell \nu_\ell$) it is
possible to have both like-sign (LS) and opposite-sign (OS) $\mu$-$e$
final states: $\mu^\pm e^\pm$ and $\mu^\pm e^\mp$.
%%%%%%
In order to be able to reconstruct the top, decaying via DFV processes, we will look only for final
states containing an electron, a muon
and a jet, identified as coming from a hadronic tau decay ($\tau$-jet),
a jet identified as coming from a $b$ quark ($b$-jet),
and a certain number of jets produced by the $u_i$ ($u_i=$ $u$, $c$) hadronization
and from the initial (ISR) and final state radiation (FSR).
Because of the presence of neutrinos in the final states, we will also
require to have missing transverse energy in the detector.
The experimental signature adopted, in the present analysis, is then the following tri-lepton
final state:
$e \, \mu^\pm \, \tau_h^\mp  \, + \, \met \, + \, b \, + \,$ jets.
%%%%%%%%%%%%%%%%%%%%%%%%%%%%%%%%%%%%%%%%%%%%%%%%%%%%%%%%%%%%%%%%%%%%%%%%%%%%%%%%%%%%%%%%%%%%%%%%%%%

%% prop: Tevatron out of reach
%%%%%%%%%%%%%%%%%%%%%%%%%%%%%%%%%%%%%%%%%%%%%%%%%%%%%%%%%%%%%%%%%%%%%%%%%%%%%%%%%%%%%%%%%%
As the cross section, for the channel under study, ranges, for $\sqrt{s}=$ $1.96$ TeV,
between $1.4$ fb and $0.1$ fb, for a Higgs mass between $114.4$ and $165$ GeV$/c^2$,
Higgs medited double flavor violating top quark decays are out of reach of Tevatron
experiments.
%%%%%%%%%%%%%%%%%%%%%%%%%%%%%%%%%%%%%%%%%%%%%%%%%%%%%%%%%%%%%%%%%%%%%%%%%%%%%%%%%%%%%%%%%%

%%% Analysis CUTS: Event selection
%%%%%%%%%%%%%%%%%%%%%%%%%%%%%%%%%%%%%%%%%%%%%%%%%%%%%%%%%%%%%%%%%%%%%%%%%%%
Higgs mediated DFV top quark decays and SM backgound events have been generated
using the $\Pythia$ $8.108$ Montecarlo program. Initial and final-state
radiation, hadronization and decay have been included. Jets have been reconstructed
using the $\Getjet$ cone algorithm and a toy Montecarlo simulation of the detector effect
efficiencies have been included in the calculations.
 Events have been selected by applying the following set of cuts.
%%%%%%%%%%%%%%%%%%%%%%%%%%%%%%%%%%%%%%%%%%%%%%%%%%%%%%%%%%%%%%%%%%%%%%%%%%%
%%%
%%%
%%%
%%%
%%% MUON
%%%%%%%%%%%%%%%%%%%%%%%%%%%%%%%%%%%%%%%%%%%%%%%%%%%%%%%%%%%%%%%%%%%%%%%%%%%
The presence of a central, isolated, high-$P_T$ muon
with $P^\mu_T>$ $20$ GeV/$c$ and $|\eta_\mu |<$ $2.4$ ~\cite{MET}.
%%%%%%%%%%%%%%%%%%%%%%%%%%%%%%%%%%%%%%%%%%%%%%%%%%%%%%%%%%%%%%%%%%%%%%%%%%%
%%%
%%%
%%% TAU
%%%%%%%%%%%%%%%%%%%%%%%%%%%%%%%%%%%%%%%%%%%%%%%%%%%%%%%%%%%%%%%%%%%%%%%%%%%
The presence of a central, isolated, high-$P_T$, hadronically decayed, $\tau$,
tagged as a $\tau$-jet ~\cite{ATLASCMS} with
$P^\tau_T>$ $20$ GeV/$c$ and $|\eta_\tau |<$ $2.4$.
%%%
%%%
%%%  Jets
%%%%%%%%%%%%%%%%%%%%%%%%%%%%%%%%%%%%%%%%%%%%%%%%%%%%%%%%%%%%%%%%%%%%%%%%%%%
Jets are reconstructed with a cone size of $R=\sqrt{(\Delta \eta)^2 + (\Delta \phi)^2} = 0.5$
and the events are required to have at least two jets with a trasverse energy of
$E^{\rm jet}_T > 30~{\rm GeV}$ within a fiducial volume of
$|\eta_{\rm jet}| <2.4$ (here the $\tau_h$s are not counted as jets).
%%%
%%%
%%% MET
%%%%%%%%%%%%%%%%%%%%%%%%%%%%%%%%%%%%%%%%%%%%%%%%%%%%%%%%%%%%%%%%%%%%%%%%%%%
Because of the presence of neutrinos in the final state, the missing transverse energy is required to be:
$\met >$ $60$ GeV and the $\mu$ and the $\tau$ are asked to have opposite sign.
%%%
%%%
%%%  b-jet
%%%%%%%%%%%%%%%%%%%%%%%%%%%%%%%%%%%%%%%%%%%%%%%%%%%%%%%%%%%%%%%%%%%%%%%%%%%
The events are then required to have a jet tagged as a $b$ and to have
%%%%%%%%%%%%%%%%%%%%%%%%%%%%%%%%%%%%%%%%%%%%%%%%%%%%%%%%%%%%%%%%%%%%%%%%%%%
%%%
%%%
%%%  ELECTRON
%%%%%%%%%%%%%%%%%%%%%%%%%%%%%%%%%%%%%%%%%%%%%%%%%%%%%%%%%%%%%%%%%%%%%%%%%%%
a central, isolated, high-$P_T$ electron with $P^e_T>$ $20$ GeV/$c$ and $|\eta_e |<$ $2.4$.
%%%%%%%%%%%%%%%%%%%%%%%%%%%%%%%%%%%%%%%%%%%%%%%%%%%%%%%%%%%%%%%%%%%%%%%%%%%
%%%
%%%
A further cut, on the reconstructed transverse mass $M_T(\mu,\tau,j)$ of the
muon, the tau and the jet, with the highest $P_T$, not tagged as a $b$-jet,
is applied in order do discrimate the signal from the SM backgrounds.
%%%%%%%%%%%%%%%%%%%%%%%%%%%%%%%%%%%%%%%%%%%%%%%%%%%%%%%%%%%%%%%%%%%%%%%%%%%

%% Simulation tools
%%%%%%%%%%%%%%%%%%%%%%%%%%%%%%%%%%%%%%%%%%%%%%%%%%%%%%%%%%%%%%%%%%%%%%%%%%%%%%%%%%%%%%%%%%%%%%%
%%%%%%%%%%%%%%%%%%%%%%%%%%%%%%%%%%%%%%%%%%%%%%%%%%%%%%%%%%%%%%%%%%%%%%%%%%%%%%%%%%%%%%%%%%%%%%%
%% Backgrouds
%%%%%%%%%%%%%%%%%%%%%%%%%%%%%%%%%%%%%%%%%%%%%%%%%%%%%%%%%%%%%%%%%%%%%%%%%%%%%%%%%%%%%%%%%%%%%%%
The SM backgrounds, considered in the present analysis, include  $t \bar t$ events,
$ZZ+$jets, $WW+$jets, $ZW+$jets and $Z+\gamma$ with the $\gamma$
that later converts in the detector giving a lepton pair.

The Large Hadron Collider, will start to operate soon delivering
the first proton proton collisions. The luminosity will be ramping
from initial $3 \times 10^{28}$ to about $2 \times 10^{31}$
cm$^{-2}$ s$^{-1}$, within first months of LHC commissioning. In
the second phase of LHC operation (within $1$-$2$ years), the
luminosity should achieve the designed value of $10^{34}$
cm$^{-2}$ s$^{-1}$ with a bunch crossing time of $25$ ns. %%
In total, an integrated luminosity of $300$ fb$^{-1}$ should be
collected, within the first $5$ years of LHC operation and about
$700$ fb$^{-1}$ before the phase two of the upgrade. In total, the
integrated luminosity delivered, in course of LHC and SLHC
running, is expected to reach a value of about $3000$
fb$^{-1}$~\cite{Nash}.
%%%%
In the scenario considered here, it will be possible, by using a
total integrated luminosity of $3000~\mathrm{fb}^{-1}$ of data, to rule
out, at 95 \% C.L., Double Flavor Violating top quark decays,
mediated by a Higgs, with a mass up to $155$ GeV/$c^2$ or discover
this process for a Higgs up to $147$ GeV/$c^2$.

%%%%%%%%%%%%%%%%%%%%%%%%%%%%%%%%%%%%%%%%%%%%%%%%%%%%%%%%%%%%%%%%%%%%%%%%%%%%%%%%%%%%
\section{Conclusions}
%%%%%%%%%%%%%%%%%%%%%%%%%%%%%%%%%%%%%%%%%%%%%%%%%%%%%%%%%%%%%%%%%%%%%%%%%%%%%%%%%%%%
%%
\label{c}The quark top and the Higgs boson, would constitute the
heaviest particles living at the Fermi scale. Therefore, these
particles would play an important role in flavor violating
transitions, as the mass seems to be intimately related with this
class of physics effects. It is thus expected that the large
masses of these particles tend to favor flavor violating
transitions both in the quark sector and in the lepton sector.
Apart from the large masses of these particles, the prospects of
detecting this class of new physics effects are reinforced by the
very peculiar dynamic behavior of the top quark and the strong
evidence of nonzero neutrino mass that naturally leads to lepton
flavor violating transitions mediated by the Higgs boson. Under
these circumstances, Higgs-mediated double flavor violating
transitions of the top quark could be at the reach of the next
colliders. In this paper, this possibility has been explored in a
model-independent manner using the effective Lagrangian approach.
A Yukawa sector that includes all the $SU_L(2)\times
U_Y(1)$-invariant operators of up to dimension six was proposed
and used to construct the most general renormalizable flavor
violating $q_iq_jH$ and $Hl_il_j$ vertices. Low-energy data were
used to constraint the $tu_iH$ and $H\tau \mu$ couplings and then
used to predict the double flavor violating decay $t\to u_i\tau
\mu$. It was found that this decay has a branching ratio of order
of $10^{-4}-10^{-5}$ for a relatively light Higgs boson with mass
in the range $115$ GeV$/c^2$$<m_H<2m_W$. Such kind of decays are
out of the reach of Tevatron experiments. Considering that LHC
will operate as a veritable top quark factory, by studying the
following final state: $e \, \mu^\pm \, \tau_h^\mp  \, + \, \met
\, + \, b \, + \,$ jets and by using a total integrated luminosity
of $\rm 3000~fb^{-1}$, it will be possible to rule out, at 95\%
C.L., DFV top quark decays up to a Higgs mass of $155$ GeV/$c^2$
or discover such a process up to a Higgs mass of $147$ GeV/$c^2$.

%%%%%%%%%%%%%%%%%%%%%%%%%%%%%%%%%%%%%%%%%%%%%%%%%%%%%%%%%%%%%%%%%%%%%%%%%%%%%%%%%%%%
\ack{We thank CONACYT (M\'exico), the HELEN ALFA-EC program
and the Universit\'a di Cassino (Italy) for their financial support.\\[0.5cm]}
%%%%%%%%%%%%%%%%%%%%%%%%%%%%%%%%%%%%%%%%%%%%%%%%%%%%%%%%%%%%%%%%%%%%%%%%%%%%%%%%%%%%

%%%%%%%%%%%%%%%%%%%%%%%%%%%%%%%%%%%%%%%%%%%%%%%%%%%%%%%%%%%%%%%%%%%%%%%%%%%%%%%

\begin{thebibliography}{99}
\bibitem{MET} We use a ($z,\phi, \theta$) coordinate system where the $z$-axis
    is in the direction of the proton beam, and $\phi$ and $\theta$ are the azimuthal
    and polar angles respectively.  The pseudo-rapidity, $\eta$, is defined as
    $-\ln(\tan\frac{\theta}{2})$.  The transverse momentum of a charged particle is
    $P_T = P\sin\theta$, where $P$ represents the measured momentum of the charged-particle
    track.  The analogous quantity using calorimeter energies, defined as $E_T = E\sin\theta$
    is called transverse energy. The missing transverse energy is defined as
    $\met = -\mid\sum_i E_T^i\hat{n}_i\mid$ where $E_T^i$ is the
    magnitude of the transverse energy contained in each calorimeter
    tower $i$ in the pseudo-rapidity region $\mid\eta\mid <3.6$ and
    $\hat{n}_i$ is the direction unit vector of the tower in the
    plane transverse to the beam direction.
\end{thebibliography}
\end{document}